\documentclass[english]{nature}

\usepackage[colorlinks,
            linkcolor=red,
            hypertexnames=false,
            anchorcolor=blue,
            citecolor=blue,
            pdftitle={Streamlined Optical Training},
            pdfauthor={Ziao WANG},
            bookmarksopen=true,
            ]{hyperref}
\usepackage{amsmath}
\usepackage{amsfonts}
\usepackage{amssymb}
\usepackage{graphicx}
\usepackage{dcolumn}
\usepackage{bm}
\usepackage[font={sf, footnotesize},labelfont=bf]{caption}
\usepackage{multirow}    
\usepackage{arydshln}
\graphicspath{ {figures/} }
\usepackage{array}
\usepackage{soul}
\usepackage{booktabs,caption}
\usepackage[flushleft]{threeparttable}

\usepackage{caption}
\usepackage{bookmark}
\renewcommand{\part}[1]{%
  \refstepcounter{part}
  \addcontentsline{toc}{part}{#1}
}

\title{Streamlined optical training of large-scale modern deep learning architectures with direct feedback alignment}

\author{
    Ziao Wang$^{1, *}$,
    Kilian M\"uller$^{2, 3, *}$,
    Matthew Filipovich$^{2, 4}$,
    Julien Launay$^{2}$,
    Ruben Ohana$^{2, 5}$,
    Gustave Pariente$^{2}$,
    Safa Mokaadi$^{2}$,
    Charles Brossollet$^{2}$,
    Fabien Moreau$^{2}$,
    Alessandro Cappelli$^{2}$,
    Iacopo Poli$^{2}$,
    Igor Carron$^{2}$,
    Laurent Daudet$^{2}$,
    Florent Krzakala$^{6}$,
    and Sylvain Gigan$^{1, \dagger}$
}

\begin{document}
\maketitle
\part{Main text}
\begin{affiliations}
    \item Laboratoire Kastler Brossel, École Normale Supérieure - Université PSL, Sorbonne Université, Collège de France, CNRS, UMR 8552, Paris, France.
    \item LightOn, 2 rue de la Bourse, 75002 Paris, France.
    \item Welinq, 14 rue Jean Mac\'{e}, 75011 Paris, France.
    \item Clarendon Laboratory, University of Oxford, Parks Road, OX1 3PU, Oxford, United Kingdom.
    \item Center for Computational Mathematics, Flatiron Institute, New York, USA.
    \item École Polytechnique Fédérale de Lausanne (EPFL), Information, Learning and Physics lab, CH-1015 Lausanne, Switzerland.

$^{*}$ These authors contributed equally to the work\\
$^{\dagger}$ Email: sylvain.gigan@lkb.ens.fr
\end{affiliations}
\begin{abstract}
Modern deep learning relies nearly exclusively on dedicated electronic hardware accelerators. Photonic approaches, with low consumption and high operation speed, are increasingly considered for inference but, to date, remain mostly limited to relatively basic tasks. Simultaneously, the problem of training deep and complex neural networks, overwhelmingly performed through backpropagation, remains a significant limitation to the size and, consequently, the performance of current architectures and a major compute and energy bottleneck. Here, we experimentally implement a versatile and scalable training algorithm, called direct feedback alignment, on a hybrid electronic-photonic platform. An optical processing unit performs large-scale random matrix multiplications, which is the central operation of this algorithm, at speeds up to 1500 TeraOPS under 30~Watts of power. We perform optical training of modern deep learning architectures, including Transformers, with more than 1B parameters, and obtain good performances on language, vision, and diffusion-based generative tasks. We study the scaling of the training time, and demonstrate a potential advantage of our hybrid opto-electronic approach for ultra-deep and wide neural networks, thus opening a promising route to sustain the exponential growth of modern artificial intelligence beyond traditional von Neumann approaches.
\end{abstract}

\section{Introduction}\label{sec1}

After decades of dominance by the Central Processing Unit (CPU) for computing all kinds of tasks, the emergence of deep learning is today a major driving force behind the development of specialized hardware. Important examples are the evolution of GPUs and the development of TPUs, both containing small but numerous processing cores, allowing them to leverage parallelization in tasks like vector-matrix multiplications that are central to today's deep learning algorithms. All these computing architectures are built upon digital hardware that follows the von Neumann paradigm~\cite{von1993first}, and the performances are fundamentally intertwined with the hardware and algorithms~\cite{hooker2021hardware}.

Optics is a promising alternative computing architecture: Except under special circumstances, light propagation is linear and can be described by a matrix that connects the input and the output fields. In other words, the propagation of light solves a vector-matrix multiplication, and it does so entirely passively, and wholly in parallel. The latter point suggests that the equivalent of the computational complexity of a vector-matrix multiplication of $O(N^2)$ on traditional hardware is reduced to $O(1)$. Practically, the finite communication bandwidth between a computer and the optical processor imposes a scaling of $O(N)$ to the speed at which data can be processed, dictated by the size of the input and output vectors. These favorable properties and the importance of this mathematical operation in data science and machine learning are the reasons for the continuous research in optical computing. Despite these promises, optical computing today remains limited to relatively small toy tasks. It therefore does not fully take advantage of this favorable scaling and lags in the current race for more efficient hardware for deep learning.

Whatever their physical implementation, most artificial neural networks (ANNs) are to date trained with end-to-end back-propagation (BP) of the error~\cite{linnainmaa1970representation, werbos1982applications, lecun2015deep}, even when implemented on physical hardware~\cite{wright2022deep}. An error is derived from the network prediction (e.g., from a supervised classification task or an unsupervised generative reconstruction task) and is used as a feedback signal to obtain parameter updates by inverting forward computations through the chain rule of derivatives. Despite decades of continuous improvement, BP bears some limitations. It is fundamentally sequential and enforces \emph{backward locking}~\cite{jaderberg2017decoupled}: a given layer may only be updated if the subsequent layer has already completed both its forward and backward pass. This hampers the simple and efficient parallelization of extreme-scale models. Even with model, data, and pipeline parallelism schemes~\cite{shoeybi2019megatron, rasley2020deepspeed}, GPU throughput during training remains bound by data transfer rates~\cite{ivanov2020data}. Alternatives to BP have, however, mainly been studied under the guise of \emph{biological realism}~\cite{lillicrap2020backpropagation}, seeking to alleviate issues such as the weight transport problem~\cite{grossberg1987competitive, lillicrap2016random}. Local learning schemes with more practical considerations have been proposed but still remain of marginal use~\cite{nokland2019training, laskin2020parallel}.

In contrast to the sequential layer updates of BP, direct feedback alignment (DFA)~\cite{nokland2016direct} uses a random projection of the error as a direct feedback signal for each layer, converting the backward pass into an entirely parallel process. DFA enables learning by approximating BP updates through a process known as alignment: forward weights eventually learn to approximate a configuration that makes the randomized feedback useful~\cite{refinetti2021align}. Among alternative training methods, DFA stands out in two ways: (1) it scales to modern deep learning tasks and architectures~\cite{launay2020direct}, which is rarely the case in alternative training methods~\cite{bartunov2018assessing}; (2) it places a single operation (a random projection) at the center-stage of training. In previous numerical studies, we have established best practices for using DFA~\cite{launay2019principled} and shown that DFA can be applied to modern deep learning tasks~\cite{filipovich2022scaling}. We have also theoretically explored the limitations of DFA's numerical implementation, convolutional layers for instance, when compared to backpropagation~\cite{refinetti2021align}, while these limitations can be quickly alleviated by modifying the vanilla algorithm~\cite{han2019efficient}.

In this work, we leverage the fact that large-scale random projections can naturally be performed optically by exploiting multiple scattering of light~\cite{saade2016random,brossollet2021lighton}. We therefore use an Optical Processing Unit (OPU) as a hardware accelerator that is perfectly matched to the DFA algorithm, as it improves the computational complexity scaling of the single operation that is its major computational bottleneck. Preliminary studies to implement DFA with optics hardware have so far focused on small-scale proof-of-concept and basic networks~\cite{launay2020hardware, filipovich2022silicon}. In the present work, we show that this combination of hardware and algorithm can train multiple types of modern digital ANN architectures, including Transformers and fully connected deep neural nets, and that it can scale to more than one billion parameters. This scale is, to our knowledge, by far a record among all non-conventional hardware training methods. The trained models demonstrate comparable final performances across diverse tasks, from language models to complex climate projections.  Finally, we study the scaling performance in terms of training time, showing that our optical training approach may provide a significant speedup for future ultra-deep and ultra-large networks. 

\section{Artificial Neural Network Training on an Optical Processor}\label{sec2}

\begin{figure*}[!htp]
  \centering
  \includegraphics[width=0.85\linewidth]{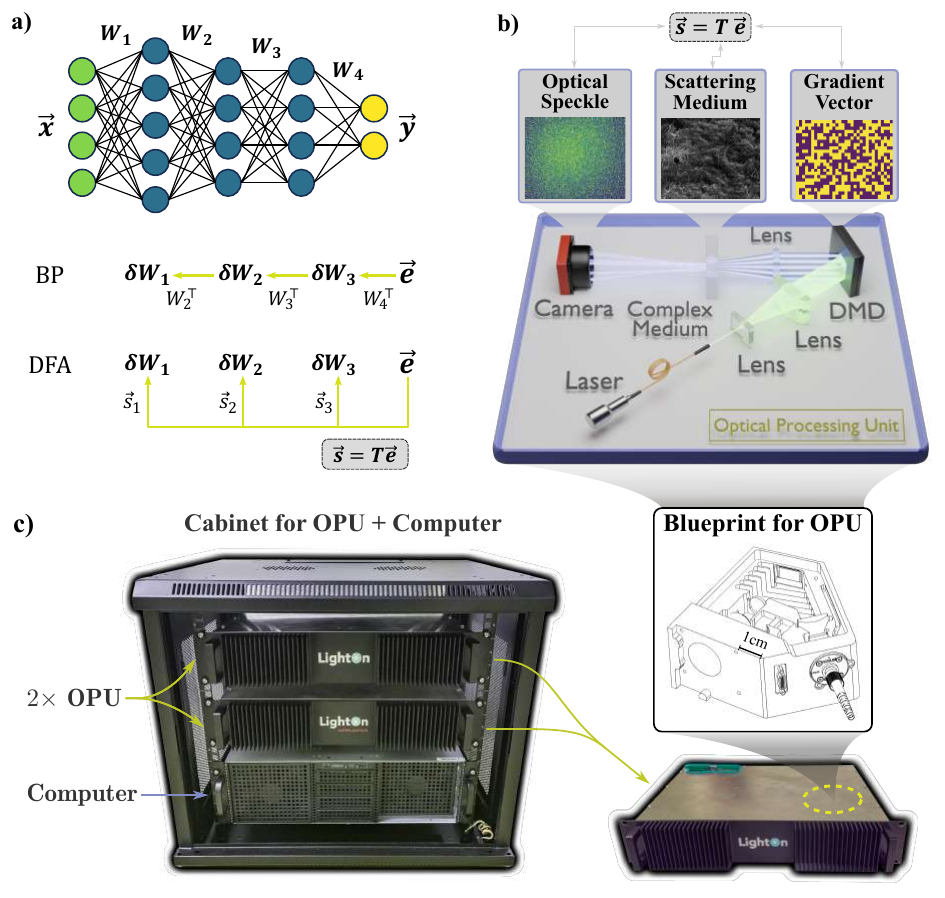}
  \caption{\textbf{Overview of the direct feedback alignment (DFA) algorithm and Optical Processing Unit (OPU).} 
    \textbf{a}, Concept of the direct feedback error propagation. Back-propagation (BP) transmits the error $\vec{e}$ sequentially from the final to the first layer, while DFA distributes error signals in parallel via random projections. 
    \textbf{b}, Illustration of the OPU. Coherent laser light illuminates a DMD, then propagates through a strongly scattering medium before being captured by a camera.  The error vector $\vec{e}$ is ternarized and encoded as binary pixels on the DMD, subsequently propagating through the complex diffusive medium to effectively perform $\mathbf{T}\vec{e}$, where $\mathbf{T}$ is a fixed, large, complex Gaussian random matrix. While the camera records an intensity pattern  $|\mathbf{T}\vec{e}|^2$ that depends non-linearly on $\vec{e}$, linear random projections can be recovered thanks to the encoding strategy (see ref.~\cite{ohana2023linear} and SI Note~\ref{sec:dfaodfa}). 
    \textbf{c}, Rack-mount cabinet for OPUs and computers. A custom enclosure houses OPUs and a computer for simultaneous operations, designed as a plug-and-play solution. Blueprint inset shows the OPU's internal layout with a $1\text{cm}$ scale. The OPU interfaces with the computer through Python libraries and is compatible with NumPy and PyTorch (see SI Note~\ref{sec:expexp}). 
    }
  \label{fig:optics}
\end{figure*}

Our OPU (see ref.~\cite{saade2016random, brossollet2021lighton} and SI Note~\ref{sec:expexp}) is based on free space optics, with the light propagating in three-dimensional space, which allows it to make the best use of the parallelism and scaling of optical computation. We use the fact that a complex medium, in which light is randomly and coherently scattered multiple times, naturally leads to a transmission matrix with random entries drawn from a normal distribution~\cite{popoff2010measuring}. Such matrices are surprisingly versatile: the field of Randomized Numerical Linear Algebra (RNLA), for example, exploits their properties to tackle high-dimensional problems~\cite{drineas2016randnla, mahoney2011randomized, martinsson2020randomized}. We have previously explored how our OPU is a natural hardware match to these RNLA algorithms~\cite{hesslow2021photonic}, and showed that it can perform different ML tasks~\cite{cappelli2021ropust, cappelli2022adversarial, ohana2021photonic, rafayelyan2020large}.  Our OPU can handle input and output vectors with dimensions up to $\sim\!1 \times 10^6$ and $\sim\!2 \times 10^6$, respectively. At this resolution, the maximum frequency is $340$ Hz, corresponding to a 1500 TeraOPS overall performance using less than $30$~Watts of power. For smaller tasks, frame rate and performance adjust accordingly (see SI Note~\ref{sec:expspeed}).

A schematic representation of the OPU is shown in Fig.~\ref{fig:optics}: A Digital Micromirror Device (DMD) modulates the incoming light field and, thus imprints the input vector $\vec{x}$ of the calculation onto the laser beam. DMDs allow the change of this input vector at kHz rates but only permit binary amplitude modulation. That is, the input vector $\vec{x}$ can only contain the elements $0$ and $1$. The complex medium gives rise to a random transmission matrix $\mathbf{T}$. Finally, the camera captures the intensity $|\vec{y}|^2 = |\mathbf{T}\vec{x}|^2$ of the output field $\vec{y}$. We further exploit two additional features. We first implement a method to obtain linear random projections from such intensity measurements without holography~\cite{ohana2023linear} (see SI Note~\ref{sec:dfaodfa}). Secondly, we ternarize the input data into vectors containing only $\{-1, 0, 1\}$, which comes at no discernible performance loss when using the DFA algorithm. Separate random projections of the positive and negative parts are combined in a post-processing step. In our hybrid training approach we are at liberty to train some parts of an ANN via ODFA, while others can be trained via BP.

\section{Training a generative Transformer on natural language processing} \label{sec3}

\begin{figure*}[!htp]
  \centering
  \includegraphics[width=0.85\linewidth]{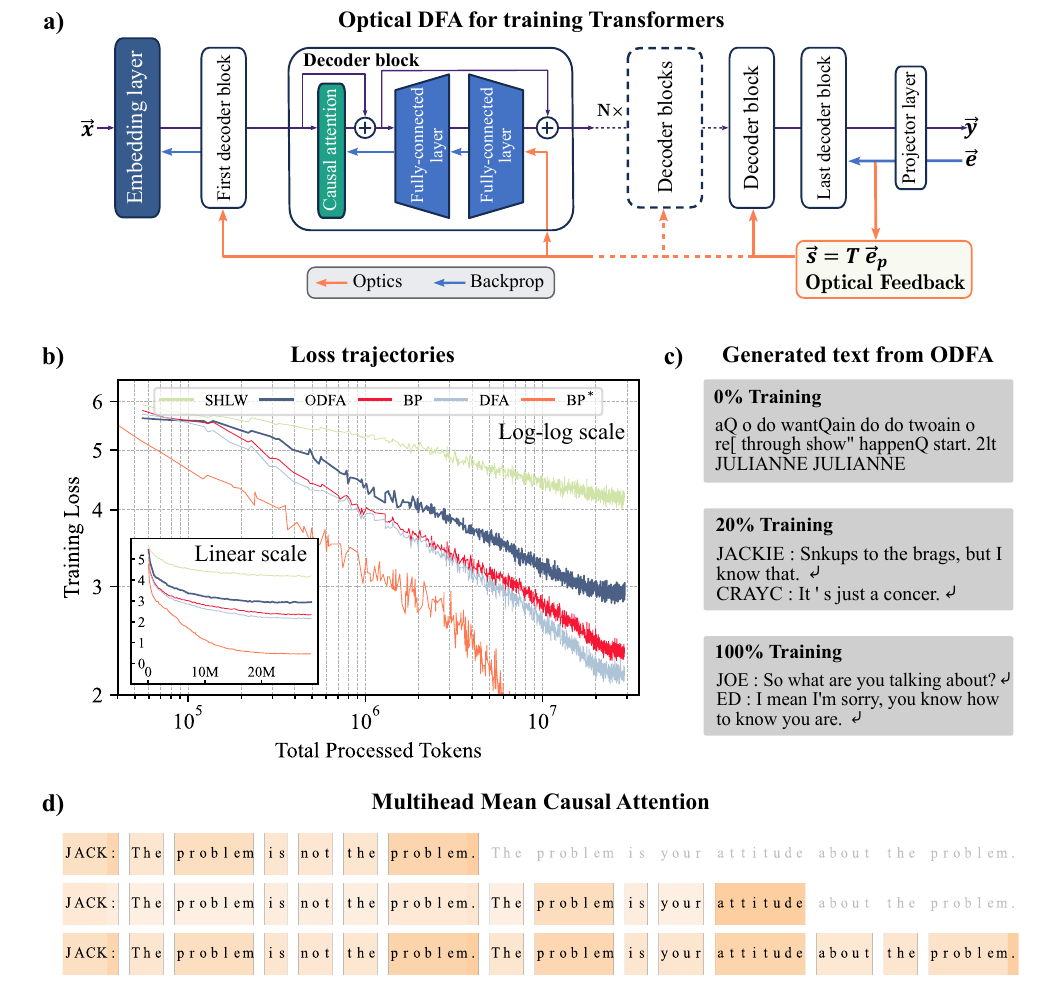}
   \caption{
   \textbf{Optical training of a generative Transformer for Movie-Dialogs dataset.} 
       \textbf{a}, Schematic representation of our Optical DFA (ODFA) training algorithm. The gradient vector $\vec{e}_p$ from the last layer is multiplied by a random matrix $\mathbf{T}$ and sent to each decoder block other than the last one in parallel, while local backprop is applied within each block, with no gradient communication among blocks.
       \textbf{b}, Loss trajectories of the generative Transformers trained using various methods. All methods employed the same architecture, but the first four utilized the same ODFA-adopted training configuration. BP$^*$ applied a distinct one where backpropagation can reach the best performance. SHLW trained the last decoder layer solely using backpropagation, with the other parameters frozen (see SI Note~\ref{sec:llmconfig}).
       \textbf{c}, Examples of text generated by the ODFA-trained Transformer at $0\%$, $20\%$, and $100\%$ training stages respectively. A bent arrow represents the generated newline token.
       \textbf{d} Mean causal attention for a given prompt (at $100\%$ training). We used causal attention, meaning that the tokens in the sequence only incorporate themselves and the previous tokens. The translucent words indicate that the Transformer has not yet processed the tokens. Shades of orange represent the attention weight on certain words.
    }
  \label{fig:llm}
\end{figure*}

To demonstrate the feasibility of the proposed hybrid approach, we employed optical DFA (ODFA, see SI Note~\ref{sec:dfaodfa}) to train a Transformer with $1.07$ billion parameters (comparable to GPT2~\cite{radford2019language}). Among the parameters of the decoder blocks of the Transformer architecture, 330M parameters directly received the ODFA signals as the gradients (see SI Note~\ref{sec:llmconfig}). We trained this Transformer on the Cornell Movie-Dialogs Corpus~\cite{danescu2011chameleons}, which consists of the characters’ names and their conversations. We divided the corpus of texts into $\sim\! 1.2\times 10^6$ sub-word tokens (i.e., the dataset size) which are composed of $1016$ unique tokens (i.e., the vocabulary size). Given a sequence of tokens as the input (context), a generative Transformer for natural language processing determines the next token by calculating the probability of each token in the vocabulary. While the inference of Transformers with optics~\cite{anderson2023optical} and end-to-end optical networks~\cite{spall2025training} have been investigated, large-scale optical training remains unexplored.

Figure~\ref{fig:llm}(a) depicts the architecture of our trained generative language Transformer and its integration with ODFA. Figure~\ref{fig:llm}(b) shows the loss trajectories to compare different training methods. Aside from ODFA, DFA, and BP, we added one Shallow Training (SHLW) baseline to validate that ODFA results in a better performance than assigning random static weights to the inner blocks. One can observe that over two epochs, all the training methods enable Transformers to learn the vocabulary's distribution, as shown by their downward-trending loss curves. We can also observe that the final losses of BP and DFA are at the same level (see SI Note~\ref{sec:llmconfig} for further discussion) and that ODFA, while it learns somewhat more slowly, still shows a consistent learning pace throughout the training with the same configuration. As a concrete example, the ODFA-trained Transformer was prompted with, ``JACK: The problem is not the problem. The problem is your attitude about the problem". The generated text (Fig.~\ref{fig:llm}(c)) was observed at different stages of the training process. Initially, the Transformer generated nonsensical combinations of sub-words and punctuations. As the training progressed, the Transformer learned to generate text in a more conversational format, with capital character names at the start of each line, one newline command at the end, and increasingly meaningful words and phrases. 

For this proof of concept, we adjusted the Transformer parameters to minimize the number of optical projections. In particular, we used a much shorter context length ($24$ tokens instead of the usual $\sim\!1000$ tokens) and a larger embedding dimension ($2040$ instead of the conventional $\sim\!500$) (see SI Note~\ref{sec:llmresult}). Increasing the embedding dimension did not lead to a longer processing time for the OPU.  As a result, the quality of the text generated by the ODFA method remains behind the state of the art. This low quality is principally due to the limited context length, which limits the ability to manage long sentences and dialogues: the generated text tends to lack logical connections between utterances. The choice to limit context size is due to the number of optical projections scaling linearly with the number of tokens and the context size. To complete the training within a reasonable time (about $20$ hours for ODFA compared to approximately $2$ hours with conventional BP in the same configuration), we thus trained the model on a small corpus and reduced context size. Note that the scale of the random projection for language tasks (typically $10^3$ inputs to $10^3$ outputs) was still limited concerning the potential dimension of the OPU ($10^6$ inputs and outputs), resulting in a comparatively slower rate of random projection on the OPU relative to the GPU at this scale. Therefore, this example should be considered as a demonstration of the ability of ODFA to train a large-scale Transformer architecture, and not a showcase of its sheer performance or speed. That being said, the large dimension of the OPU is potentially well adapted for frontier models, whose vocabulary sizes are often in the range of $10^5$, inner model dimensions typically reach $10^4$ for dense models, and effectively $10^5$ for Mixture-of-Experts like GPT-4~\cite{chowdhery2023palm}.

\section{Vision Transformer and deep neural networks on climate task} \label{sec4}

\begin{figure*}[!htp]
  \centering
  \includegraphics[width=0.9\linewidth]{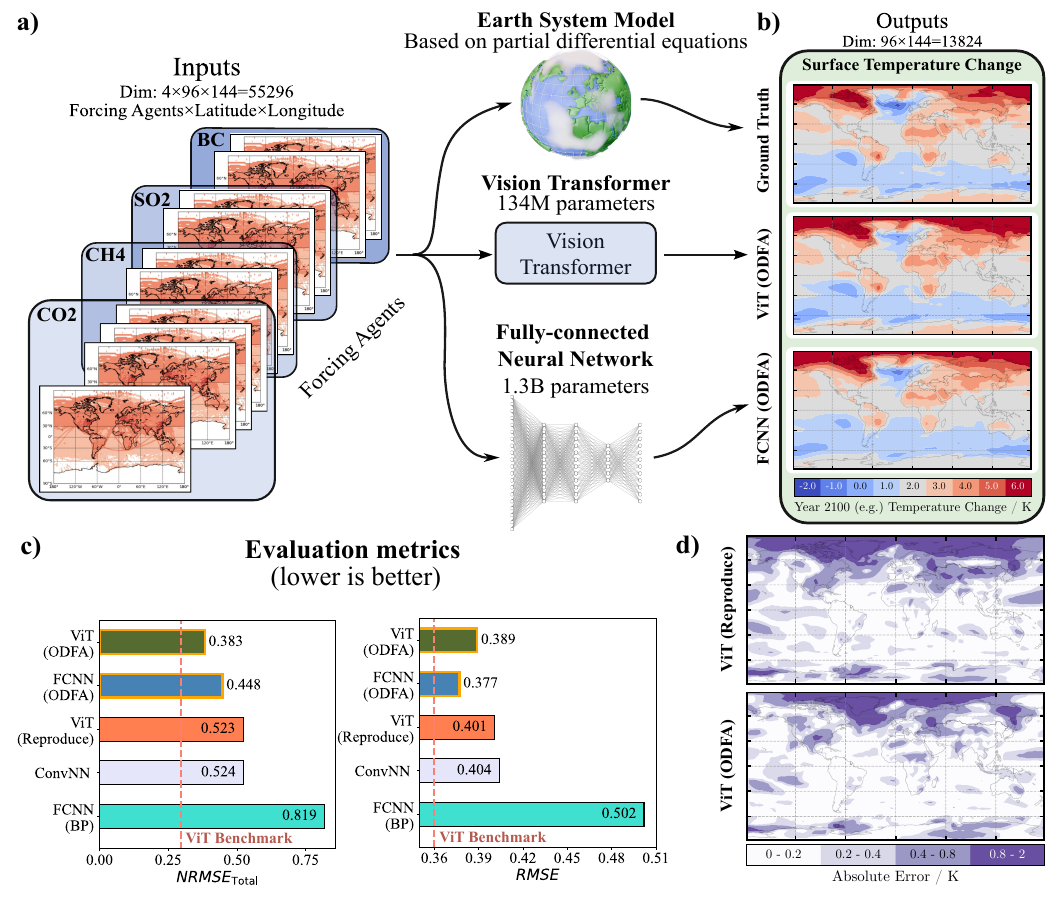}
  \caption{\textbf{ODFA training of ViT and FCNN on the high-dimensional climate projection dataset.} 
    \textbf{a}, Schematic illustration of BP/ODFA-trained ViT/FCNN on the climate projection dataset. The inputs are the global distributions of four forcing factors (e.g., carbon dioxide) from 2015 to 2100, with a dimension of $55.3$k. The outputs are the global distributions of surface air temperature at specific years, with a dimension of $13.8$k. The dataset contains both historical recordings for past years and simulation results based on Earth system models for future years. One ViT with 134M parameters and one FCNN with 1.3B parameters were trained both using BP and ODFA. 
    \textbf{b}, Ground truth and predictions of ODFA-trained ViT/FCNN on the temperature change in Year 2100. 
    \textbf{c}, Performances of BP/ODFA-trained models over two RMSE-based metrics. The BP results of \textit{ViT(Reproduce)} were obtained using the same configuration as \textit{ViT(ODFA)}, and the results of \textit{ConvNN} are from~\cite{Watson-Parris2022} (see SI Note~\ref{sec:vitmetric}). The ViT benchmark (dashed red line) corresponds to the optimal training of the ViT using a much larger dataset~\cite{Nguyen2023}. 
    \textbf{d}, Absolute error between the targets and the BP/ODFA-trained ViT predictions respectively. Both present a similar error level. 
  }
  \label{fig:climate}
\end{figure*}

To further test the effectiveness of ODFA in configurations that play more to its strengths, we trained a Vision Transformer (ViT), a Transformer architecture suited for computer vision tasks, on the climate projection dataset ClimateBench~\cite{Watson-Parris2022}. Given the heterogeneity of climate projection data sources and the complex nature of the input and output data, the ViT has been identified as an effective model for the task~\cite{Nguyen2023}. Here, the ViT was trained to map and predict the global distributions of four anthropogenic forcing factors to the distribution of the Earth's surface air temperature, with large input and output dimensions of $55296$ and $13824$, respectively (Fig.~\ref{fig:climate}(a)). Using the same ODFA framework in the language Transformer, we trained the ViT: 60 million of its 134 million parameters directly received the ODFA signal, while the rest relied on local backpropagation to exclusively propagate the ODFA signal within each decoder block (see SI Note~\ref{sec:vitdata}). As illustrated in Fig.~\ref{fig:climate}(b), the predictions of the ODFA-trained ViT (second row) closely resemble the ground truth distribution (first row). Based on a comparison of absolute errors between ODFA- and BP-trained ViTs under the same configuration in Fig.~\ref{fig:climate}(d), we observed both approaches remain a similar error magnitude, differing only in finer details. A quantitative evaluation in Fig.~\ref{fig:climate}(c) verifies that utilizing ODFA to train ViTs performs comparably to BP. Notably, for this specific task the ODFA method overcame the large performance gap between electronics and optics that we observed in the language task. Indeed, our ODFA-trained ViT is close to the state-of-the-art ViT benchmark model~\cite{Nguyen2023}.

Next, to highlight the capability of ODFA to train different types of architectures, we trained a fully connected neural network (FCNN) on the same dataset. The architecture of the FCNN in this study consists of $4$ layers, with the number of nodes decreasing in the deeper layers and a total of 1.3 billion parameters. For consistency, the last layer was again trained by BP. During training, the gradients from the last layer were optically projected onto the ODFA signal and subsequently sent in parallel to the other layers. Here, 1.03 billion parameters directly received the ODFA signal.

Figure~\ref{fig:climate}(c) summarizes the evaluation metrics of BP and ODFA. Here, the BP-trained FCNN performed much worse than the ODFA-trained one. The gain in performance can be attributed in part to the training configurations during the backward pass. For such a wide dimensional layer, the gradients in the backward pass require additional task-specific normalization and careful adjustments of the step size and batch size. In practice, Layer Normalization is sometimes used in the forward pass to re-center and re-scale the gradient in the backward pass~\cite{xu2019understanding}. However, the optical projected signal is naturally normalized throughout the laser and camera calibration, ensuring the camera is not over-exposed naturally leads to a normalized ODFA signal. Therefore, for FCNN, an ODFA-adopted training configuration can lead to an uncharacteristically bad result for using BP.

To further showcase ODFA's applicability beyond language and vision tasks, we explore diffusion-based generative methods, especially the Diffusion Transformer (DiT)~\cite{peebles2023scalable}, a modern architecture designed for iterative noise-to-image synthesis. Adopting the same methodology as in language and vision Transformers, we integrate ODFA into DiT through minimal architectural modifications (see SI Note~\ref{sec:ditarch}). Our first evaluation on MNIST (see SI Note~\ref{sec:ditmnist}) shows ODFA's capacity to guide diffusion models, achieving convergence comparable to BP in this low-complexity domain, and matching the digit-generation quality. We then extend ODFA training for DiT to a substantially more challenging task, Animal Faces (AFHQv2)~\cite{choi2020stargan}, which demands handling greater color diversity, structural complexity, and stylistic variation. From our results (see SI Note~\ref{sec:ditanimal}), despite some artifacts, ODFA maintains stable training, producing animal faces with recognizable species identity. Although still at an early training stage, these results validate ODFA's cross-modal adaptability, even in advanced diffusion-based frameworks without compromising core denoising functionality.

\section{Scaling towards extreme-scale models} \label{sec5}

\begin{figure*}[!htp]
  \centering{
  \includegraphics[width=0.92\linewidth]{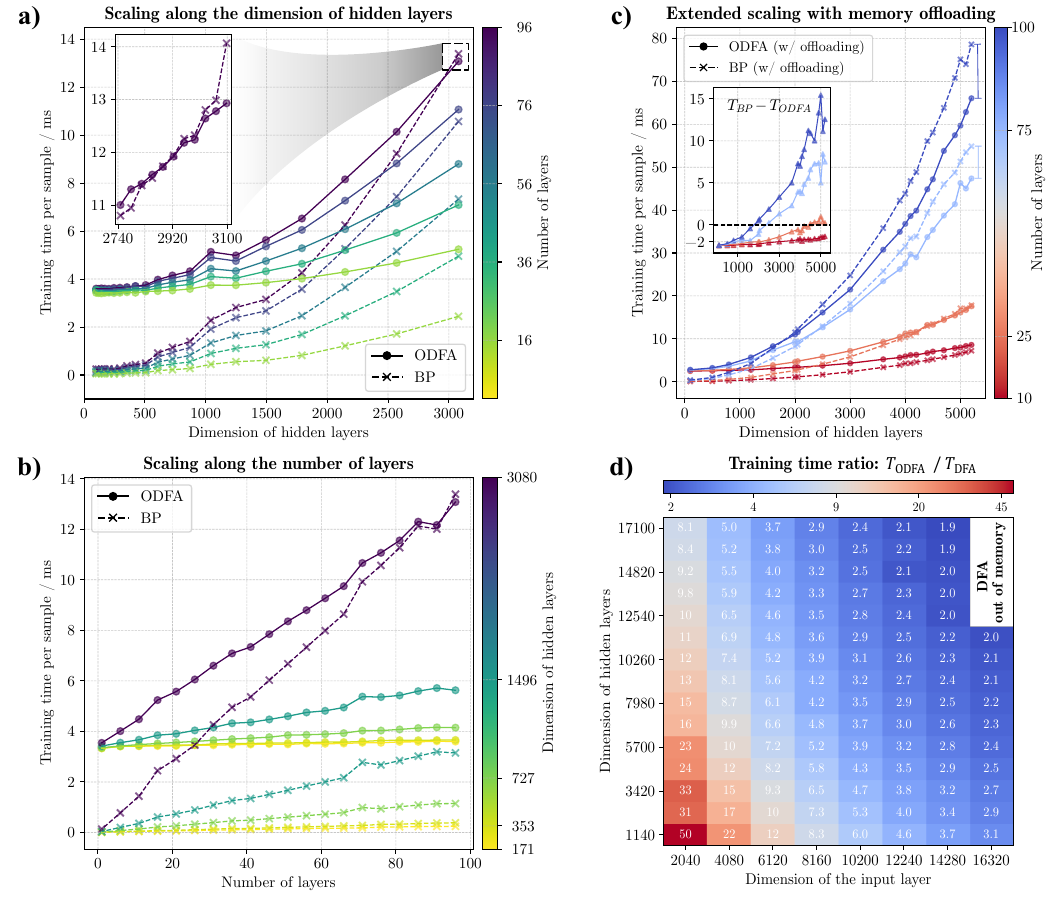}}
  \caption{\textbf{Scaling of the training time per sample for ODFA-trained extreme-scale FCNNs.} 
    \textbf{a}, Training time of FCNNs versus hidden layer size ($100$ to $3080$ neurons per hidden layer). Dot-solid lines represent training using ODFA; cross-dashed lines are for BP. Both follow the expected quadratic behavior. Curve colors indicate hidden layer count: yellow for shallow, purple for deep FCNNs. At the rightmost configuration ($96$ layers, $3080$ neurons each), BP ($13.39$ ms/sample) is slower than ODFA ($13.09$ ms/sample), reaching the GPU's memory limit. Inset highlights more data around this configuration. 
    \textbf{b}, Training time versus number of layers ($1$ to $96$). Curve colors indicate hidden layer size: yellow for narrow, purple for wide FCNNs. Training time increases linearly with layer count. 
    \textbf{c}, Extended comparison using offloading technique to overcome GPU memory limitations, along hidden layer sizes ($100$ to $5200$), up to 2.7 billion parameters. Curve colors indicate hidden layer count: red for shallow, blue for deep FCNNs. Both BP and ODFA employ the same offloading strategy. ODFA presents a sustained speed advantage over BP at extra-large scales. Inset indicates the time difference. 
    \textbf{d}, Ratio of training time: ODFA (GPU-OPU) vs. DFA (GPU only). The time difference narrows at higher dimensions until the GPU memory limit is reached. 
  }
  \label{fig:scaling}
\end{figure*}

Scaling up deep learning models can boost performance~\cite{kaplan2020scaling}, but remains resource-intensive and time-consuming. Under GPU-based training compute budgets, which unify the estimations of both resource and time usage, DFA generally underperforms BP~\cite{filipovich2022scaling}. By contrast, ODFA's optical operations decouple training time from GPU-centric compute constraints. We thus study the training time of ODFA and BP across scaled FCNN with synthetic data. To compare the training time of BP and ODFA (in real-world seconds), we measured the total elapsed wall-clock time required to complete the entire training process in each case. These measurements therefore encompass all algorithmic and communication steps (see SI Note~\ref{sec:elsgpu}). As discussed below, the results summarized in Fig.~\ref{fig:scaling} show that DFA, in combination with an optical accelerator, can scale to massive sizes. They also demonstrate that the scaling pre-factors are smaller for ODFA than for BP and therefore hint at a speed advantage compared to pure in-silico GPU training for the largest NNs.

Specifically, in Fig.~\ref{fig:scaling}(a), we varied the dimension of the hidden layers (number of neurons in each hidden layer) while leaving the numbers of layers fixed, corresponding to a quadratically increasing number of parameters (weights) of neural networks with the number of neurons. When the dimension of the hidden layers is small, BP can process training quickly, in microseconds per sample, for any number of layers. We observe the expected quadratic growth of the training time as the dimension grows. For ODFA, the frame rate of the DMD and the camera both set a minimum training time of a few ms, even for a minuscule model. While the ODFA part is independent of the dimension, the update of the parameters and the forward pass remain quadratic. However, the advantage of ODFA becomes obvious in the extreme-scale region, where the calculation of the updated weights via BP becomes dominant. In Fig.~\ref{fig:scaling}(b), for a fixed dimension of the hidden layers, we can observe the linear scaling of the training time for BP with the number of layers commensurate with the scaling of the number of parameters of the network. As expected from the scaling of parameters, where increasing the number of layers leads to a linearly increasing number of parameters, the training time for both methods rises with a nearly constant slope. However, the optical method always has a smaller slope again thanks to the $O(1)$ scaling of the random projections with the number of parameters. Ultimately, we can reach a regime where the training speed of ODFA surpasses BP in our case (for $96$ layers and $3080$ neurons in each layer). To verify that this advantage holds and the trend is sustained when pushing to extra-large scale, we leveraged a standard offloading technique to bypass the GPU memory limits (see SI Note~\ref{sec:elsol}). With an extended scaling in Fig.~\ref{fig:scaling}(c), ODFA maintained a notable and robust shorter training time over BP, with models up to 2.7 billion parameters ($100$ layers, $5200$ neurons each). This confirms ODFA's scalability and robustness towards the extra-large scale.

We now focus on the training-time reduction, arising both from the algorithm and the optical hardware integration. The DFA algorithm bypasses the protracted gradient propagation through layers in the backward pass, thereby increasing the parallelization. The expected reduction of the training time for certain epochs is around $24\%$ at massive scales. However, this reduction does not exclusively stem from the algorithm. Considering the maximum computing dimension for our optical hardware ($10^6\times 10^6$) and the dimension of the current gradient projection ($10^3\times 10^3$), the optical hardware is far from working at its full potential. The projection dimension only depends on the output layer dimension and the largest hidden layer dimension. Fig.~\ref{fig:scaling}(d) shows the decreasing ratio of the DFA training time with and without the OPU for different projection dimensions. The ratio decreases from $50$ at the scale of $10^3\times 10^3$ to 1.9 at $10^4\times 10^4$. Beyond this point, training a larger architecture with BP or DFA will exceed the storage capacity of our GPU, NVIDIA A6000, with 32GB of storage. However, since the coefficients of the large random matrix are stored optically, ODFA can scale beyond DFA's dimensions without much speed penalty (see SI Note~\ref{sec:elsdfa} for further demonstration).

\section{Discussion} \label{sec6}

We have presented a symbiotic pairing of non-standard hardware (optics) and training algorithm (DFA) to train large-scale artificial neural networks. We have demonstrated the versatility of this system by training GPT-like language Transformer, Vision Transformers, Diffusion Transformer and fully connected neural networks on language, vision, and diffusion-based tasks. Further, we have shown its scalability by training up to 1B parameters, to our awareness, a record for any training scheme using non-conventional hardware.

The fundamental motivation for this work is the promise that pairings between specialized hardware and algorithms can provide pathways outside of the existing ``hardware lottery"~\cite{hooker2021hardware}. The compelling advantage of our system is its ability to execute the central calculation of the DFA algorithm (a random vector-matrix multiplication) optically for a very large dimension $N$ and within a favorable $O(N)$ time complexity and energy consumption scaling. We have demonstrated this scaling and compared it to the $O(N^2)$ scaling of traditional hardware (GPU) by adapting neural network architectures to the specific strengths of our system. A single one of our current Optical Processing Units (OPU), which is based on free space optics, is capable of processing input and output vectors with $N \sim 10^6$ entries, which is an order of magnitude larger than the requirement of current frontier models. With available off-the-shelf components (SLMs and cameras), $N$ could be increased by another order of magnitude. These properties of optical computation (scaling and large $N$), combined with the highly parallelizable DFA algorithm, hint at the possibility of training extremely large models with a substantial computational advantage.  Our approach may therefore provide a sustainable pathway for the training of more capable models, simply through an increase in the number of parameters and the adaptation of the ANN's architecture to the strengths of the ODFA system.

\section*{References}
\bibliographystyle{naturemag}
\bibliography{ref}

\renewcommand\thefigure{S\arabic{figure}}
\renewcommand\thesection{Supplementary Note \arabic{section}.}
\renewcommand\thesubsection{\arabic{section}.\arabic{subsection}}
\setcounter{section}{0}
\setcounter{figure}{0}

\newpage
\part{SI Note}

{\spacing{1}\setlength{\parskip}{2pt}{\Large\bfseries\noindent\sloppy \textsf{\centering{Supplementary Materials for:}\\ Streamlined optical training of large-scale modern deep learning architectures with direct feedback alignment}} \par
\noindent
    Ziao Wang$^{1, *}$,
    Kilian M\"uller$^{2, 3, *}$,
    Matthew Filipovich$^{2, 4}$,
    Julien Launay$^{2}$,
    Ruben Ohana$^{2, 5}$,
    Gustave Pariente$^{2}$,
    Safa Mokaadi$^{2}$,
    Charles Brossollet$^{2}$,
    Fabien Moreau$^{2}$,
    Alessandro Cappelli$^{2}$,
    Iacopo Poli$^{2}$,
    Igor Carron$^{2}$,
    Laurent Daudet$^{2}$,
    Florent Krzakala$^{6}$,
    and Sylvain Gigan$^{1, \dagger}$
}

\noindent{\textbf{\large{Contents}}}\\
\textbf{Supplementary Note 1. Direct feedback alignment and optical training}\\
\ref{sec:dfadfa} Direct feedback alignment\\
\ref{sec:dfaodfa} Ternarized encoding and quantization of the optical training \\
\textbf{Supplementary Note 2. Experimental details and noise impact}\\
\ref{sec:expexp} Experimental setup\\
\ref{sec:expspeed} Operation speed and energy consumption\\
\ref{sec:expstab} System stability\\
\ref{sec:expnoise} Impact of noise in the optical training (Toy model)\\
\textbf{Supplementary Note 3. Large language model trained by ODFA}\\
\ref{sec:llmdata} Movie-Dialogs dataset and the preprocessing for the ODFA-trained language Transformer\\
\ref{sec:llmtoken} Tokenization and configuration of the language Transformer architecture\\
\ref{sec:llmconfig} ODFA for Transformers and training configurations\\
\ref{sec:llmresult} Performance of the language Transformers trained by different methods\\
\ref{sec:llmnoise} Impact of noise in the optical training (Large-scale Transformer)\\
\textbf{Supplementary Note 4. ODFA for vision tasks}\\
\ref{sec:vitdata} Preprocessing for Climate projection task and configurations\\
\ref{sec:vitmetric} Evaluations of the ViT trained by different methods\\
\textbf{Supplementary Note 5. ODFA on diffusion models}\\
\ref{sec:ditarch} ODFA for Diffusion Transformers (DiT)\\
\ref{sec:ditmnist} DiT trained with ODFA on MNIST dataset\\
\ref{sec:ditanimal} DiT trained with ODFA on Animal Faces dataset\\
\textbf{Supplementary Note 6. Further details about scaling toward extreme-scale models}\\
\ref{sec:elsgpu} Training time measurement workflow with GPU and OPU\\
\ref{sec:elsol} Extra-large scaling with memory offloading\\
\ref{sec:elsdfa} Extended comparison of ODFA and DFA

\noindent{\textbf{Supplementary Figures:}}\\
Figure \ref{fig:tdfa}. Performance of DFA and TDFA on MNIST\\
Figure \ref{fig:stability}. Speckle feature stability of the experimental setup\\
Figure \ref{fig:noise}. Simulation of ODFA with TM perturbation noise on MNIST\\
Figure \ref{fig:dfa2noise}. Simulation of ODFA with measurement noise and transmission matrix drifting\\
Figure \ref{fig:tokenize}. Performance of the Transformer trained by ODFA with different tokenizers\\
Figure \ref{fig:components}. Performance of the 1B-parameter Transformer with various ODFA scenarios\\
Figure \ref{fig:moreepochs}. Training curves with more epochs\\
Figure \ref{fig:contextsize}. ODFA training with varying context sizes\\
Figure \ref{fig:testloss}. Training curves and test loss on 1M- vs. 2M-Token Datasets\\
Figure \ref{fig:llmnoise}. Noise impact on the 1B-parameter Transformer in the optical training\\
Figure \ref{fig:vitfcnn}. FCNN trained by ODFA on Climate projection task\\
Figure \ref{fig:vitbitbetter}. Grid-by-grid comparison of ViT trained by different methods\\
Figure \ref{fig:metrics}. Performances of BP- or ODFA-trained models over four RMSE-based metrics\\
Figure \ref{fig:ditarch}. Diffusion Transformer (DiT) architecture and integration of ODFA\\
Figure \ref{fig:ditmnloss}. Training curves of a Diffusion Transformer with BP, DFA, and ODFA\\
Figure \ref{fig:ditmngen}. Class-conditional MNIST samples from DiT-B/2 trained by different methods\\
Figure \ref{fig:ditmnchain}. Reverse diffusion chains of DiT-B/2 trained by different methods on MNIST\\
Figure \ref{fig:ditafloss}. Training curves of a Diffusion Transformer on Animal Faces dataset\\
Figure \ref{fig:ditafgenchain}. Samples of DiT-B/2 trained by different methods on Animal Faces dataset\\
Figure \ref{fig:elsgpufull}. Training time of FCNNs versus hidden layer size across different layer counts\\
Figure \ref{fig:elsolfull}. Training time with memory offload of increasing depth and width\\
Figure \ref{fig:elstmfull}. Extended time difference with memory offload\\
Figure \ref{fig:elsdfaol}. ODFA vs. DFA training time ratio with offloading

\section*{Overarching Claim}
We reiterate here that every mention of ODFA through the main text and this supplementary note refers to a real experiment involving our optical setup, OPU.

\section{Direct feedback alignment and optical training} \label{sec:dfa}
\subsection{Direct feedback alignment} \label{sec:dfadfa}
Training a deep neural network on a supervised dataset always requires updating its weights given only the prediction of the network and the desired output. Conventionally, back-propagation (BP) is virtually always used to accomplish this by sequentially updating all the layers. The direct feedback alignment (DFA) algorithm is an appealing alternative way to update the weights by propagating the error directly from the output to each hidden layer of the network in parallel using random feedback signal $s_l$.

Consider a neural network $f$ parameterized with $L$ layers, where each layer is denoted by the index $l$, with $l\in 1, 2, \dots, L$.
Each layer is associated with an activation function $g$, a bias vector $\underline{b}^{(l)}\in\mathbb{R}^{d_l}$ and a weight matrix $\underline{\underline{W}}^{(l)}\in\mathbb{R}^{d_{l}\times d_{l-1}}$, where $d_l$ refers to the number of neurons in layer $l$.
For each layer $l$, given the activation $\underline{a}^{(l-1)}\in\mathbb{R}^{d_{l-1}}$ of the previous layer, the activation of layer $l$ reads
\begin{equation}
    \underline{h}^{(l)} = \underline{\underline{W}}^{(l)}\underline{a}^{(l-1)} + \underline{b}^{(l)},\quad
    \underline{a}^{(l)} = g\left(\underline{h}^{(l)}\right),
\end{equation}
where $\underline{a}^{(0)} = \underline{x}$ is the input data for the network.
Given the loss function $\mathcal{L}$ and the error signal $\underline{e}\equiv \frac{\partial\mathcal{L}}{\partial \underline{h}^{(L)}}$, the update of the last layer of weights is calculated as
\begin{equation}
     \underline{\underline{\Delta W}}^{(l)} = -\eta \underline{\delta h}^{(l)} \left(\underline{a}^{(l-1)}\right)^{\top},\quad 
     \underline{\Delta b}^{(l)} = -\eta \underline{\delta h}^{(l)},
\end{equation}
for a learning rate $\eta$, where $\underline{\delta h}^{(L)} = \underline{e}$ is the error signal for the weights of the last layer.
For BP, the factors $\underline{\delta h}^{(l)}_{\mathrm{BP}}$ of other layer weights are defined sequentially as,
\begin{equation}
    \underline{\delta h}^{(l)}_{\mathrm{BP}} = \frac{\partial \mathcal{L}}{\partial \underline{h}^{(l)}} = 
    \left( \left( \underline{\underline{W}}^{(l+1)}\right)^{\top}\underline{\delta h}^{(l+1)}_{\mathrm{BP}} \right) \odot 
    g^{\prime}\left(  \underline{h}^{(l)} \right)
\end{equation}
with $\odot$ denoting the Hadamard product.

For DFA, the factor $\underline{\delta h}^{(l)}_{\mathrm{DFA}}$ is calculated as
\begin{equation}
    \label{eq:dfa}
    \underline{\delta h}^{(l)}_{\mathrm{DFA}} = \underline{s}^{(l)} \odot g^{\prime}\left(  \underline{h}^{(l)} \right),\quad
    \underline{s}^{(l)} = \underline{\underline{T}}^{(l)}\underline{e},
\end{equation}
where the transpose of the network weights $\left(\underline{\underline{W}}^{(l+1)}\right)^{\top}$ is replaced by a fixed random matrix $\underline{\underline{T}}^{(l)}\in \mathbb{R}^{d_l \times d_{L}}$, and thus DFA allows to update the parameters of all layers in parallel.

DFA, a variant of the Feedback Alignment (FA) algorithm~\cite{lillicrap2016random}, thus provides an alternative, gradient-free method to train neural networks and other deep learning architectures using a direct feedback path from the error to each layer.
The only requirement of DFA is access to the error vector and a fixed random matrix $\underline{\underline{T}}$. The matrices $\underline{\underline{T}}^{(l)}$ can be sub-matrices of $\underline{\underline{T}}$ and thus share elements without noticeable impact on the training~\cite{launay2019principled}. This is very advantageous since, practically, we use sub-matrices of the one optical random matrix that we have at our disposal. Importantly, prior knowledge of the matrix $\underline{\underline{T}}$ is not necessary for the training.
This is a merit of our optical training scheme using a random transmission matrix as $\underline{\underline{T}}$ to update the weights of large-scale models. If prior knowledge of the elements $T_{ij}$ was required, then for optical training we would need to measure the exact transmission matrix $\underline{\underline{T}}$.
These prerequisite measurements would be time-consuming. 
Meanwhile, the required measurements scale with the dimension of input and output, which could obstruct leveraging optical training on large-scale datasets or models.
Since precise calibration of $T_{ij}$ is not required for  DFA, optical training using DFA is then further accessible.

\subsection{Ternarized encoding and quantization of the optical training} \label{sec:dfaodfa}
DFA and BP involve massive matrix-vector multiplications, as shown in Eq.~\ref{eq:dfa}, which can be moved from an electronic platform to an optical layer. Here, we perform the random projection $\underline{\underline{T}}^{(l)}\underline{e}$ on our Optical Processing Unit (OPU). The basic idea is to replace the random matrix $\underline{\underline{T}}$ with the transmission matrix of a complex medium, whose random elements are drawn from a normal distribution.
The error $\underline{e}$ is first collected at the output of the network after the forward pass and ternarized such that it can be displayed on the binary pixels of a digital micromirror device (DMD). More specifically, the elements of the normalized error vector $\underline{e} = \{e_1, e_2, \dots, e_{d_L}\}^{\top}$ are mapped to the values $\{-1;0;1\}$ using a threshold $\pm t$, resulting in two binary vectors $\underline{e}^+$ and $\underline{e}^-$ with elements,
\begin{equation}
    e_i^+ = \left\{
    \begin{aligned}
        & 0, && \mathrm{if}\; e_i < t, \\
        & 1, && \mathrm{else}
    \end{aligned}\right.
    ,\quad
    e_i^- = \left\{
    \begin{aligned}
        & 1, && \mathrm{if}\;e_i > -t, \\
        & 0, && \mathrm{else}.
    \end{aligned}\right.
\end{equation}
Using $\underline{e}^+$ and $\underline{e}^-$ we obtain the random feedback signal via $\underline{s}^{(l)}=\underline{\underline{T}}^{(l)}\underline{e}^+ - \underline{\underline{T}}^{(l)}\underline{e}^- = \underline{\underline{T}}^{(l)}\left(\underline{e}^+ - \underline{e}^-\right)$.
This Ternarized DFA (TDFA) scheme requires two projections through the OPU for each training step but conserves a better approximation of the error signal.
To implement the random projection optically, the laser beam containing the DMD-encoded information of $\underline{e}$ propagates through a scattering medium.
The scattering medium is described by a complex-valued Gaussian distributed transmission matrix $\underline{\underline{\tilde{T}}}\in\mathbb{C}^{[\max_{1\leq i <L}(d_i)]\times d_L}$.
An intensity measurement by the camera captures the modulus-square $\left\lVert \underline{\underline{\tilde{T}}}\,\underline{e} \right\rVert^2$.
We have developed a method to obtain linear optical random projections from intensity measurements that notably don't require holographic techniques~\cite{ohana2023linear}.
Briefly, a new linear transform can be constructed via two additional random projections involving a constant anchor vector $\underline{r}\in \mathbb{R}^{d_L}$,
\begin{equation}
    \underline{s} = \underline{\underline{T}}\,\underline{e} = \frac{\left\lVert \underline{\underline{\tilde{T}}}\,\underline{r} \right\rVert^2 + \left\lVert \underline{\underline{\tilde{T}}}\,\underline{e} \right\rVert^2 - \left\lVert \underline{\underline{\tilde{T}}}\left(\underline{r}-\underline{e}\right) \right\rVert^2}{2\sqrt{\left\lVert \underline{\underline{\tilde{T}}}\,\underline{r} \right\rVert^2}},
\end{equation}
where $\underline{\underline{T}}$ refers to the linear random projection matrix that we use in our DFA experiments, and $\underline{\underline{\tilde{T}}}$ is the optical transmission matrix of the OPU, with otherwise the same previous notations.
Notably, in this case, the effective random matrix $\underline{\underline{T}}$ is no longer the same as the optical transmission matrix $\underline{\underline{\tilde{T}}}$.
This method simplifies the experimental setup and increases the stability since there is no need for interference with an external reference light beam.
Based on this formula, we can perform linear projections through a Gaussian-distributed matrix with our OPU.

\begin{figure*}[!htp]
  \centering
  \includegraphics[width=1.\linewidth]{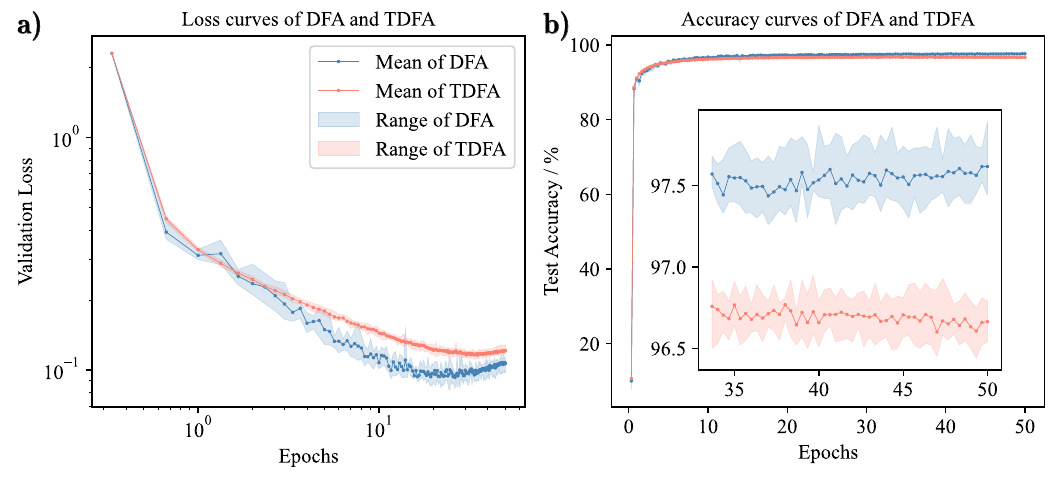}
  \caption{\textbf{Performance of DFA and TDFA on MNIST} 
    DFA and TDFA were used to train a three-layer neural network on the full MNIST dataset with the following layer dimensions: $[784, 100, 10]$. The batch size was $100$, and the model was trained over $50$ epochs. TDFA encodes the gradient vectors into three values, $[-1, 0, 1]$. The results were obtained by $20$ runs with the same configuration.
    \textbf{Left}, Validation loss along the training process by DFA and TDFA. The mean is the mean validation loss over the $20$ models with the same configuration and the same for the range. 
    \textbf{Right}, Test accuracy along the training process by DFA and TDFA. A zoom-in inset shows the test accuracy after $30$ epochs.
  }
  \label{fig:tdfa}
\end{figure*}

The quantization of error vectors has an influence on the final performance of optical training and the question of how it influences it requires further exploration.
The first practical issue is to determine the ternarization threshold $t$. This hyperparameter is in principle trainable for each element of the gradient during the training process, but this approach would lose the simplicity and adaptability of the DFA algorithm.
Therefore, in this project, we simply selected a global threshold $t$ at the beginning of the training based on the distribution of the error vector of the first batch.
To decide the global threshold $t$, we maximized the cosine similarity between the projected vector of DFA and TDFA, as,
\begin{equation}
    \arg\,\max_{0\leq t} = S_C\left(\underline{s}_{\mathrm{DFA}}(\underline{e}_0), \underline{s}_{\mathrm{TDFA}}(\underline{e}_0, t)\right)
    = \frac{\underline{s}_{\mathrm{DFA}}(\underline{e}_0) \cdot \underline{s}_{\mathrm{TDFA}}(\underline{e}_0, t)}{\left\lVert\underline{s}_{\mathrm{DFA}}(\underline{e}_0)\right\rVert  \left\lVert\underline{s}_{\mathrm{TDFA}}(\underline{e}_0, t)\right\rVert},
\end{equation}
where $\underline{e}_0$ is the gradient vector of the first training step.
Here, we selected the (normalized) cosine similarity as an objective function because the gradient vector's direction plays a role in the optimisation whereas its magnitude does not.
To compare the performance of TDFA to that of DFA we trained a two-layer neural network on the MNIST dataset (Fig.~\ref{fig:tdfa}).
One can notice that during the training process, the validation loss of TDFA consistently remained slightly higher than DFA. 
As the second practical issue, there was a significant gap in the final test accuracy between TDFA and DFA.
However, we believe that a $1\%$ gap in the test accuracy is acceptable when using TDFA, considering that we convert the continuous-valued error into ternarized vectors. 
This is because the training process can compensate for the small loss in accuracy by running for a longer period, thanks to the increased speed of the binarized-value input device.

\section{Experimental details and noise impact} \label{sec:exp}
\subsection{Experimental setup} \label{sec:expexp}
We employ a LightOn Appliance OPU to perform linear random projections. While the overall OPU is a 2U rack unit, the optical system itself (without the laser) is roughly $16\mathrm{cm}\times12\mathrm{cm}\times8.5\mathrm{cm}$.  The experimental setup includes a $532$nm continuous-wave laser a DMD for encoding the gradient vector from the last layer, a disordered scattering medium that provides the random transmission matrix of the system, and a camera to detect the speckle patterns, which are then used to calculate the linear random projection (see SI Note~\ref{sec:dfaodfa}). The optical setup communicates with the computer via PCIe through an I/O interface module that handles data transfer, parts of the data formatting, synchronization, and other related tasks.

More in detail, the laser light (Oxxius LCX-532L) emerges from a polarisation-maintaining fiber, and is collimated by a set of lenses before illuminating the DMD (Texas Instruments DLP4500 with Ajile AJD-4500-UT controller), which features $912\times 1140$ micro-mirrors. Each micro-mirror pixel encodes gradient vectors information onto the laser beam as a binary amplitude modulation by tilting between $-12^{\circ}$ or $12^{\circ}$, thereby determining whether the light at this pixel is directed towards the scattering medium, or diverted to a beam dump. The light beam carrying this spatially encoded information is focused onto the scattering medium via a lens. After having propagated through this medium, the optical signal is registered by the camera (Basler Aca2000 340KMS) containing $2048\times 1088$ pixels, where each pixel provides an $8$-bit digital output.

We note that the DMD operates at $2880$ Hz in continuous streaming mode. The camera achieves $340$ Hz when reading the full sensor, but by using the row-windowing mode, the frame rate can reach up to $5.5$ kHz. The overall OPU frame rate is constrained by the slower component, which in this project is typically the camera. However, this bottleneck could be alleviated by integrating newer, commercially available imaging technologies, such as Ximea CB019MG-LX-X8G3, that allows up to $2.28$ kHz at approximately 2 million pixels.

\subsection{Operation speed and energy consumption} \label{sec:expspeed}
In this subsection, we will first calculate the effective operation speeds of the optical setup, and then present the energy consumption of each computing part in ODFA.

\textbf{Operation speed.} To estimate the overall computing performance of the OPU, we first calculate the number of equivalent operations executed by a single optical random projection. For an input vector $\underline{e}\in\mathbb{R}^m$ and the transmission matrix $\underline{\underline{T}}\in\mathbb{R}^{n\times m}$, each pass through the OPU involves $n\times m$ multiplications and $n\times (m-1)$ additions. At the maximum resolutions for both the DMD and the camera, where $n=2048\times 1088$ and $m=912\times 1140$, the total number of operations per projection is $n(2m-1)\approx 4.6\times 10^9$. Given that the device can operate at $340$~Hz at this resolution, the maximum computing performance is approximately $1500$ TeraOPS.

Here, we present in detail the measured operation speed for each task in the main text. In the language task, the OPU itself operates at $1460$~Hz, corresponding to $6.03$~GigaOPS. In the climate projection task, the OPU operates at $1440$~Hz ($24.2$~GigaOPS) for ViT, and $1400$~Hz ($0.27$~TeraOPS) for FCNN. 

However, there are three caveats limiting direct comparison with conventional electronic computing. First, each ternarized linear random projection requires four passes through the OPU, effectively lowering the operation speed by a factor of four. Second, the OPU's input and output for ODFA in this work were restricted to ternary and 8-bit data, respectively, reducing precision. Third, the standard operation speed is calculated assuming full control of each element, which requires additional memory storage, whereas our OPU uses a naturally given random matrix that cannot be controllably changed. Consequently, the OPU’s $1500$ TeraOPS is not directly comparable to modern electronic chips due to incomplete precision and limited tunability.

\textbf{Energy consumption.} To accurately quantify the energy consumption during the training with different training methods, we employed external power meters to continuously monitor the system's total power draw. The power meter was connected directly to the socket powering the computer or the OPU. This configuration ensured that all components (GPU, CPU, OPU, fans) were included in the total power measurement. There is no other module requiring energy in the training process, as the laser, DMD, camera, control electronics, temperature controller of OPU, I/O interface module, and other non-computer electronics have been included into OPU's energy consumption. During the 2-hour measurement on training the language Transformer in the main text Fig.~\ref{fig:llm}, the ambient temperature was set at $25$ degrees. We emphasize that our measurement only captures energy usage over a fixed number of steps for each method, rather than the energy required to achieve a particular model performance improvement, arguably a more practical measure for industrial applications.

According to the measurement, the OPU draws $27.1$ Watts on average (range $26.7-27.8$W), and the computer draws $474.2$W on average (range $445.8-527.3$W) during the training. When waiting for the OPU to complete the projection at each step, the computer's idle power averaged $76.7$W ($72.2-79.0$W). Notably, total power consumption depends not only on the training method, but also the model architecture, which can vary widely. For example, to train a model ($4200$ neurons per layer, $100$ layers) in the main text Fig.~\ref{fig:scaling}(c), energy usage reached $26.1$ Joule/sample with BP, $21.6$ J/sample with DFA, and $19.6$ J/sample with ODFA, showing a modest advantage for ODFA. 

As an initial takeaway, these results suggest that ODFA can offer favorable computing performance and energy efficiency compared to standard backpropagation, but at the cost of incomplete precision, limited tunability, and moderate performance.

\subsection{System stability} \label{sec:expstab}
\begin{figure*}[!htp]
  \centering
  \includegraphics[width=1.\linewidth]{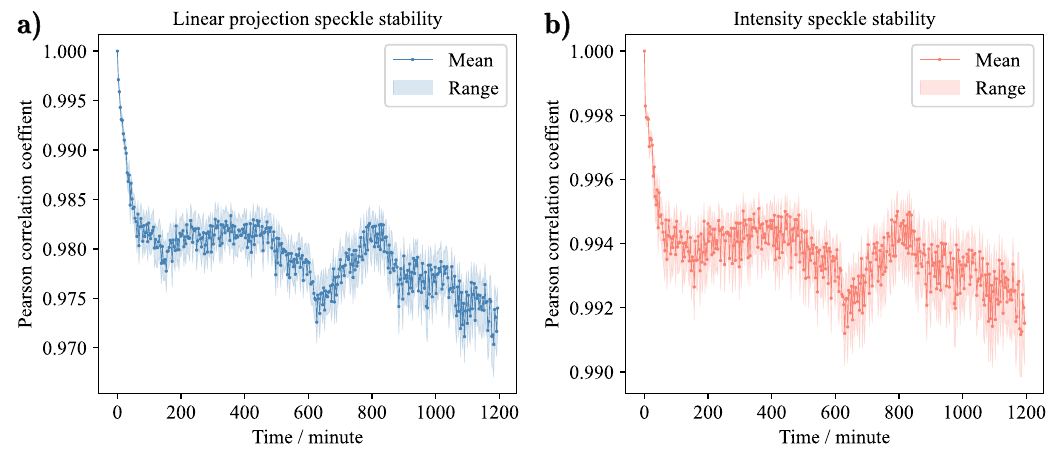}
  \caption{\textbf{Speckle feature stability.} 
    The stability of the speckle features over $1200$ minutes, within which the whole training process was finished. Here, we applied the Pearson correlation coefficient (PCC) to evaluate the stability.
    \textbf{a,} The stability of the speckle features captured by the linear projection. The PCC was calculated between the speckles at a certain time and the beginning, given the same reference input. The mean PCC and the range of the PCC were calculated over $20$ reference inputs, where each reference has $2040$ modes, the same as the number of modes required by optical training of the generative language model in the main text Fig.~\ref{fig:llm}.
    \textbf{b,} The stability of the speckle features captured by the intensity measurement. This one has a higher PCC, which means it performs more stably because it only requires one measurement per time here compared to three times for linear projection.
  }
  \label{fig:stability}
\end{figure*}
In the algorithm detailed herein, the random matrix $\underline{\underline{T}}$ is supposed to be constant throughout the training process, in alignment with the DFA design principles. 
It underscores the experimental framework's reliance on the stability of projected vectors—termed system stability—as a cornerstone for the successful optical implementation of the training algorithm. 
Figure~\ref{fig:stability} shows the empirical evaluation of this system stability via linear projections and intensity measurements. 
This evaluation involved calculating the Pearson correlation coefficient (PCC) between the initial speckle image $\underline{s}(0)$ and subsequent speckle images $\underline{s}(t)$, while the same gradient vectors $\underline{e}$ were applied.
We note that the linear features are less stable than the intensity features. We assume that this is because several intensity measurements have to be combined in order to calculate these linear features.
We can heuristically say that this level of stability is sufficient for performing ODFA in our setup. We also want to note that we observed successful training even under less stable conditions. The impact of noise and a (slowly) drifting random matrix is discussed in more detail in the next section.

\subsection{Impact of noise in the optical training (Toy model)} \label{sec:expnoise}
\begin{figure*}[!htp]
  \centering
  \includegraphics[width=1.\linewidth]{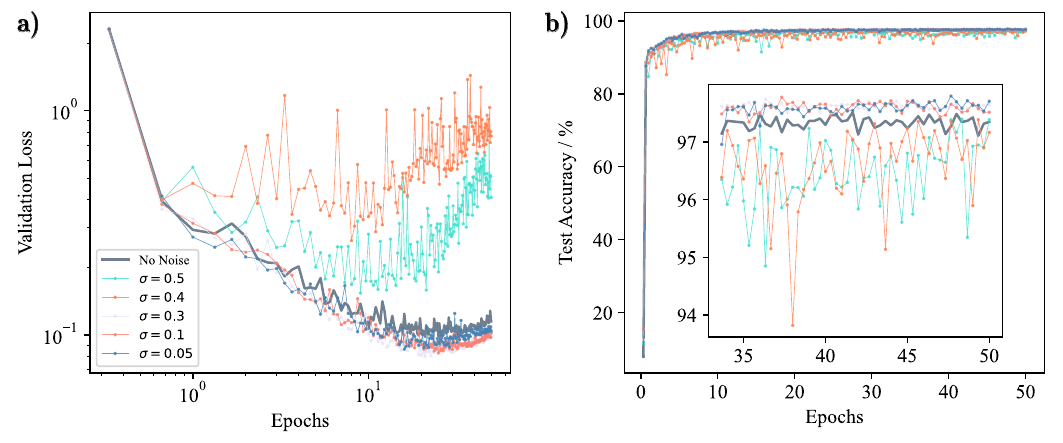}
  \caption{\textbf{Simulation of ODFA with TM perturbation noise on MNIST.} 
    Simulated ODFA was used to train a three-layer neural network on the full MNIST dataset with the following layer dimensions: $[784, 100, 10]$. The batch size was $100$, and the model was trained over $50$ epochs.
    \textbf{a,} Validation loss along the training process with different levels of noise. The elements of the random matrix used in ODFA were sampled from $\mathcal{N}(0, 1)$ with an additional noise sampled from $\mathcal{N}(0, \sigma^2)$. The random matrix was fixed during the training, while the noise was sampled at every training step. The loss curve without noise is marked with a darker color. The value of $\sigma$ varied from $0$ to $0.5$. For $\sigma=0.5$ and $\sigma=0.4$, the validation loss first decreased and then increased to around $0.6$. 
    \textbf{b,} Test accuracy along the training process with different levels of noise. A zoom-in inset shows the test accuracy after $30$ epochs. Again, the curve without any noise is marked by a darker color.
  }
  \label{fig:noise}
\end{figure*}
In the realm of optical computing, and especially for optical training, it is essential to examine the impact of noise from different parts of the experimental setup.
Although the OPU was designed to minimize perturbations induced by mechanical vibration and thermal fluctuations, the effect of the measurement noise and of the marginal instability of the system on the optical training demand further investigation.
Hereby, we implemented simulations with three kinds of noise during the training: the noise on the transmission matrix, the transmission matrix drifts, and camera noise.
The first noise corresponds to an addition to the fixed random matrix of a noise matrix, fluctuating over time (TM perturbation), resulting in $\underline{\underline{T}}^{*}(t)=\underline{\underline{T}}(0) + \underline{\underline{\Sigma}}(t)$, where $\underline{\underline{\Sigma}}(t)$ is the noise term at time $t$. 

We first define a noisy matrix $\underline{\underline{T}}^{*}(t)$ with $\underline{\underline{T}}(0)$ sampled from a Gaussian distribution $\mathcal{N}(0, 1)$ and noise $\underline{\underline{\Sigma}}$ from $\mathcal{N}(0, \sigma^2)$. Here, $t$ is a discrete time step. We then evaluate how optical DFA performs under variable levels of noise.
We train a fully connected neural network on MNIST with different levels of noise, as shown in Fig.~ \ref{fig:noise}. 
The neural network has three layers with the following layer dimensions: [784, 100, 10] and uses a vanilla SGD optimizer with a learning rate of $0.01$ and a momentum of $0.9$.
We observed that the training performance is affected by a noisy random matrix. We see that after a few epochs of continuous decrease, the validation loss starts to increase when the noise level is set to $\sigma=0.4$ and $\sigma=0.5$, and the final test accuracy for these two levels of noise is lower than the test accuracy without noise. 
However, our simulation revealed an interesting behavior: when considering a lower level of noise, such as $\sigma=0.05$, the final test accuracy and the validation loss performed better.
It is important to note that $\sigma>0.1$ corresponds to a significant amount of noise in experiments, as long as the elements of $\underline{\underline{T}}$ are sampled from $\mathcal{N}(0, 1)$.
This result shows that the ODFA algorithm can either suffer or benefit from certain levels of this kind of transmission matrix noise.

\begin{figure*}[!htp]
  \centering
  \includegraphics[width=.8\linewidth]{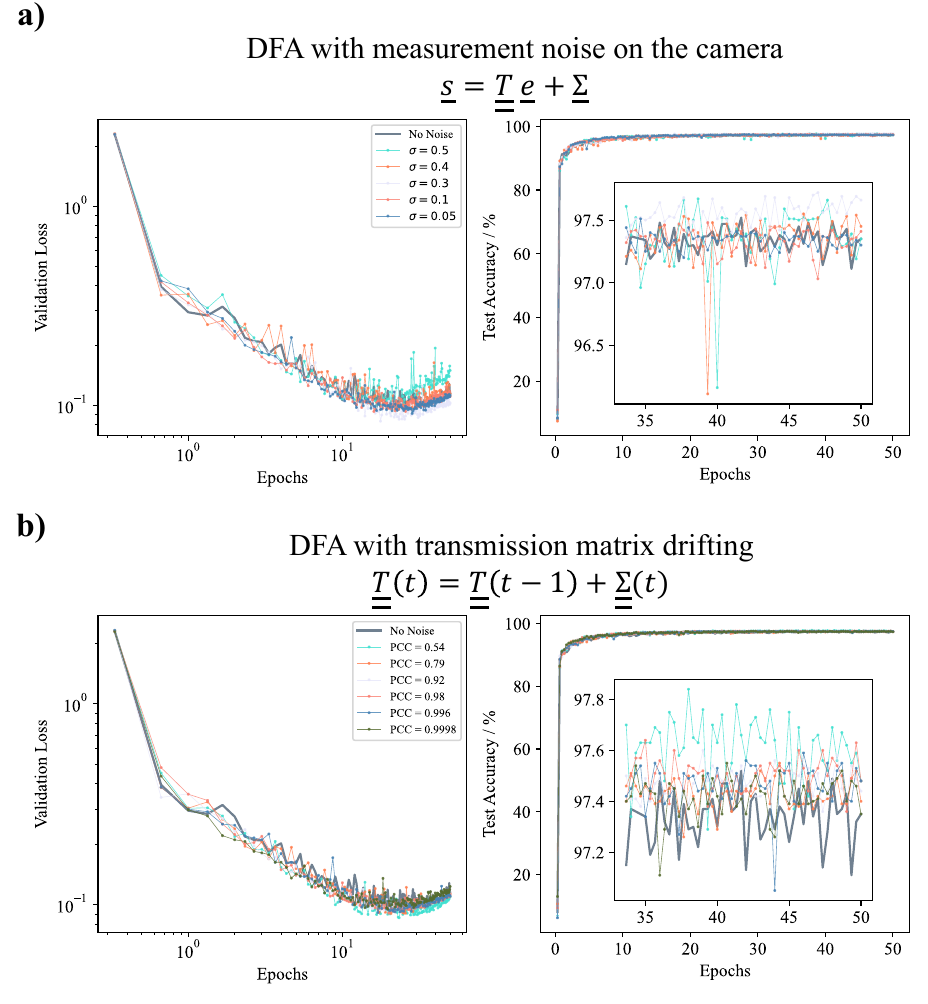}
  \caption{\textbf{Simulation of ODFA with measurement noise and transmission matrix drifting.} Similar setting as in Fig.~\ref{fig:noise}.
  \textbf{a}, Validation loss and test accuracy curves for ODFA with different levels of measurement noise. The matrix $\underline{\underline{T}}$ is fixed during the training, but the ODFA signal $\underline{s}$ suffers from a noise $\underline{\Sigma}$ whose elements are sampled from a normal distribution $\mathcal{N}(0, \sigma^2)$. The curves without any noise are marked by a darker color. A zoom-in inset again shows the test accuracy after $30$ epochs.
  \textbf{b}, Validation loss and test accuracy curves for ODFA with different levels of drifting. The matrix undergoes a drifting $\underline{\underline{T}}(t)=\underline{\underline{T}}(t-1) + \underline{\underline{\Sigma}}(t)$, where each element of the noise matrix $\underline{\underline{\Sigma}}(t)$ was sampled from a normal distribution $\mathcal{N}(0, \sigma^2)$. The drifting level is decided by the PCC between the last speckle image $\underline{s}(T_{\mathrm{end}})$ and the initial speckle image $\underline{s}(0)$.
  }
  \label{fig:dfa2noise}
\end{figure*}

We define the noisy ODFA signal as $\underline{s}(t)=\underline{\underline{T}}(0)\underline{e} + \underline{\Sigma}(t)$ for the second type of noise on the camera. 
Here, the noise $\underline{\Sigma}(t)$ is sampled from the normal distribution. 
Figure~\ref{fig:dfa2noise}(a) displays the performances under different levels of measurement noise. 
It is worth noting that the measurement noise on the camera has less impact on the final performance of the trained model than the noise on the transmission matrix $\underline{\underline{T}}$. 
The gap between ideal perfect measurement ($\sigma=0$, no noise) and extremely noisy measurement ($\sigma=0.5$) is not as significant as shown in Fig.~\ref{fig:noise}. 
Additionally, we observed that with a moderate level of noise, the final test accuracy is slightly better than without measurement noise, but not significantly so. 
Therefore, we conclude that the effect of measurement noise on the optical training performance is not significant.

Finally, to address the phenomenon of transmission matrix drifting, which constitutes the third category of noise, we delineate the transmission matrix as evolving temporally, different from the first type of noise, in which the transmission matrix fluctuates around a certain configuration.
The noise is encapsulated in the temporal drift of the transmission matrix, mathematically expressed as $\underline{\underline{T}}(t)=\underline{\underline{T}}(t-1) + \underline{\underline{\Sigma}}(t)$. 
The impact of such drifting is depicted in Fig.~\ref{fig:dfa2noise}(b). 
The quantification of noise levels is here determined by the final PCC between the initial and final speckles, which ranges from $\mathrm{PCC}=0.54$ to $\mathrm{PCC}=0.9998$. 
As a reference, our OPU system sustains a PCC of approximately $0.973$ for the linear projections over a duration of $20$ hours (i.e. $3\%$ decorrelation).
The analysis reveals that the drifting of the transmission matrix within our study does not result in a divergent validation loss throughout the training process. 
Furthermore, the test accuracy of the models remains unaffected by this specific type of noise. 
Actually, even with a PCC of $0.54$, the test accuracy remains consistent and does not seem affected significantly by the drift.
To summarize the impact of these various types of noise on the training algorithm, the optical stochastic DFA maintains the learning trajectory comparably to the ideal optical training conditions, with little apparent impact from the noise. For a further investigation on the noise impact on a more complex architecture, please refer to SI Note~\ref{sec:llmnoise}.

\section{Large language model trained by ODFA} \label{sec:llm}
\subsection{Movie-Dialogs dataset and the preprocessing for the ODFA-trained language Transformer} \label{sec:llmdata}
The language Transformer in the main text Fig.~\ref{fig:llm} has been trained by ODFA using the Cornell Movie-Dialogs Corpus~\cite{danescu2011chameleons}. 
This corpus consists of fictional conversations extracted from raw movie scripts and includes over $83,097$ conversations by $9,035$ speakers, with a total of $304,713$ utterances.
For our task, we streamlined the corpus to ensure that the training process could be completed within a reasonable timeframe. 
After consolidation, the corpus contained $6,948$ conversations with $25,867$ utterances, totaling $3,183,498$ characters. 
An example conversation from the corpus is as follows from the movie \textit{Twelve monkeys}:
\begin{quote}
RAILLY: \textit{I'll get the tickets and meet you... in the Gift Shop}.\\
COLE: \textit{Right!  You're right.  I have to fix this.}\\[1.5em]
COLE: \textit{I was here...as a kid.  I think you were here, too.  But you...looked just like you look now.}\\
RAILLY: \textit{They may be looking for us, James.  Use this.  You can fix it in the Men's Room.}
\end{quote}
Each conversation in the corpus follows a basic format where the speakers iteratively exchange several utterances. 
Each utterance is followed by the name of the speaker in capital letters and ends with a newline command. 
Additionally, there is a blank line between every two conversations.
Our first expectation for the generated text from the Transformer trained on this corpus is that it will have a similar format to the conversation and blank spaces between them.
For a text to be of higher quality, it should include conversations with two or more characters in or resembling an English language style. 
A well-generated text will have proper spelling, grammar, and logical connections, and an even better one should be with an unfolding plot like a real script.

\subsection{Tokenization and configuration of the language Transformer architecture} \label{sec:llmtoken}
\begin{figure*}[!htp]
  \centering
  \includegraphics[width=1.\linewidth]{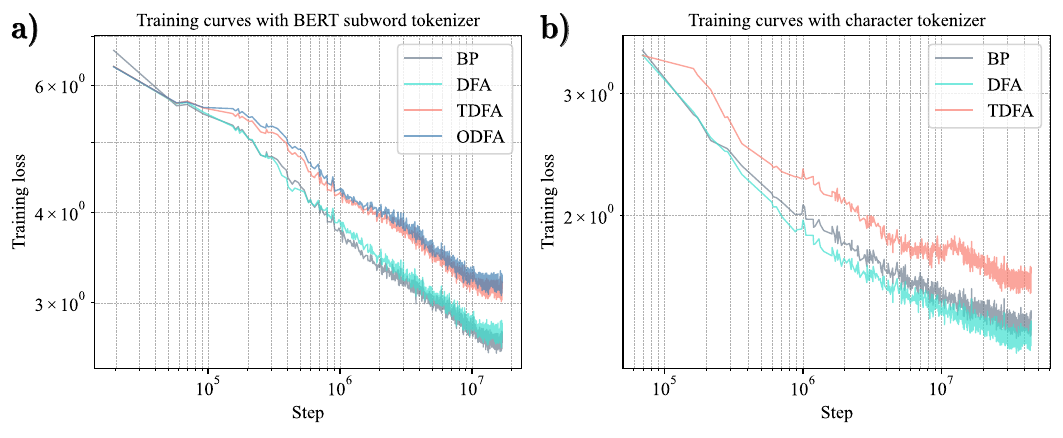}
  \caption{\textbf{Performance of the Transformer trained by ODFA with different tokenizers.}
  In this figure, the relative loss within each panel is crucial, and comparing the losses across different tokenizers would be misleading
  The Transformer was trained on a dataset using two different tokenizers: the BERT-subword tokenizer and the character tokenizer. The Transformer in the figure has the same configuration as the one in the main text Fig.~\ref{fig:llm}, and both axes in the figure are shown on a logarithmic scale.
  \textbf{Left}, Loss curves for the Transformer applied with the BERT-subword tokenizer trained by BP, DFA, TDFA, and ODFA. The size of the vocabulary list (unique tokens) used was $1016$. After tokenization, the corpus contained $1,208,114$ tokens. The optical training algorithm matched the digital algorithm version well, and the gap between the quantized version and the continuous version has already been investigated in Fig.~\ref{fig:tdfa}.
  \textbf{Right}, Loss curves for Transformer applied the character tokenizer trained by BP, DFA, and TDFA. The size of the vocabulary list is $83$ (numbers, upper and lower letters, punctuations, and symbols). The corpus had the same number of tokens before and after the tokenization, which was $3,183,498$. The final loss obtained was smaller than with the BERT-subword tokenizer as a result of a larger vocabulary.
  }
  \label{fig:tokenize}
\end{figure*}
Once the corpus is ready, the next crucial step in preparing text data for use in language Transformers is called tokenization. 
Tokenization is the process of dividing text into tokens, which can be characters, subwords, or words. 
Each token is then converted into a vector based on its index in the vocabulary and fed into the Transformer. 
Tokenization enables the Transformer to process and understand the text in a structured manner, which helps in understanding the context, semantics, and syntax of the language. 
In our case, we used two types of tokenizers, namely the natural character tokenizer and the BERT-subword tokenizer~\cite{devlin2018bert}.
The character tokenizer treats each character as a separate entity, and the size of its vocabulary depends on the number of unique characters present in the corpus. This includes numbers, letters, punctuation, and symbols. 
In contrast, the BERT-subword tokenizer uses the WordPiece algorithm~\cite{wu2016google} to generate a set number of subword tokens based on the frequency of all possible subwords. 
Due to the relatively small size of the corpus we trained on, we did not experiment with the word tokenizer, which generates a very large vocabulary that might not be useful for the Transformer to comprehend the text.

As shown in Fig.~\ref{fig:tokenize}, different tokenizers share a similar learning pace for all the training methods.
First, for both tokenizers, BP and DFA learned the corpus at a similar level and obtained a comparable final loss within each tokenizer.
The application of the subword-tokenizer did not lead to the degradation of the quantized training algorithms (TDFA and ODFA), it still learning at a similar pace to BP and DFA.
It was also observed that the relative loss gap between the continuous and the quantized algorithms maintained a consistent level.
It is noteworthy that the discrepancy in the final loss between the BERT-subword tokenizer and the character tokenizer bears no correlation with the quality of the resultant loss. 
The loss of the output, based on the probability of each token in the vocabulary being the subsequent token given a certain text sequence, scales with the size of the vocabulary. 
Consequently, the larger vocabulary associated with the BERT-subword tokenization inherently resulted in an elevated loss. To conduct a fair comparison between these two tokenizers, the generated text with the character tokenizer by ODFA is presented as follows:
\begin{quote}
    GEORLA: \textit{Youndn't give of the down.' I could you work and somethin' f}\\
    JAKE: \textit{And th school the wants a movie.}
\end{quote}
Despite the successful learning and reproduction of the basic format of the speakers and the conversations, the overall quality of the text could not match that of the text produced with the BERT-subword tokenizer. 
Upon comparison of the aforementioned text with the one in the main text Fig.~\ref{fig:llm} with the BERT-subword tokenizer, it was found that the character tokenizer resulted in the generated text with barely accurate spelling and grammar. 
The primary reason for the lack of connection between tokens can be largely attributed to the configuration we established for the Transformers.

For all the language Transformers that were trained in this subsection, the same architecture was utilized. 
The Transformer was designed in a GPT-like architecture, comprising one embedding block and several decoder blocks. 
The embedding block consisted of a token embedding layer with an embedding dimension of $2040$ in conjunction with a positional embedding layer of an identical dimension.
As for the decoder blocks, a sequence of $40$ blocks was employed, with each decoder block incorporating $10$ causal attention heads and a multilayer perception with the dimensions $[2040, 2060, 2040]$. 
In total, each Transformer contained $1,070,063,120$ trainable parameters. 

\subsection{ODFA for Transformers and training configurations} \label{sec:llmconfig}
\begin{figure*}[!htp]
  \centering
  \includegraphics[width=1.\linewidth]{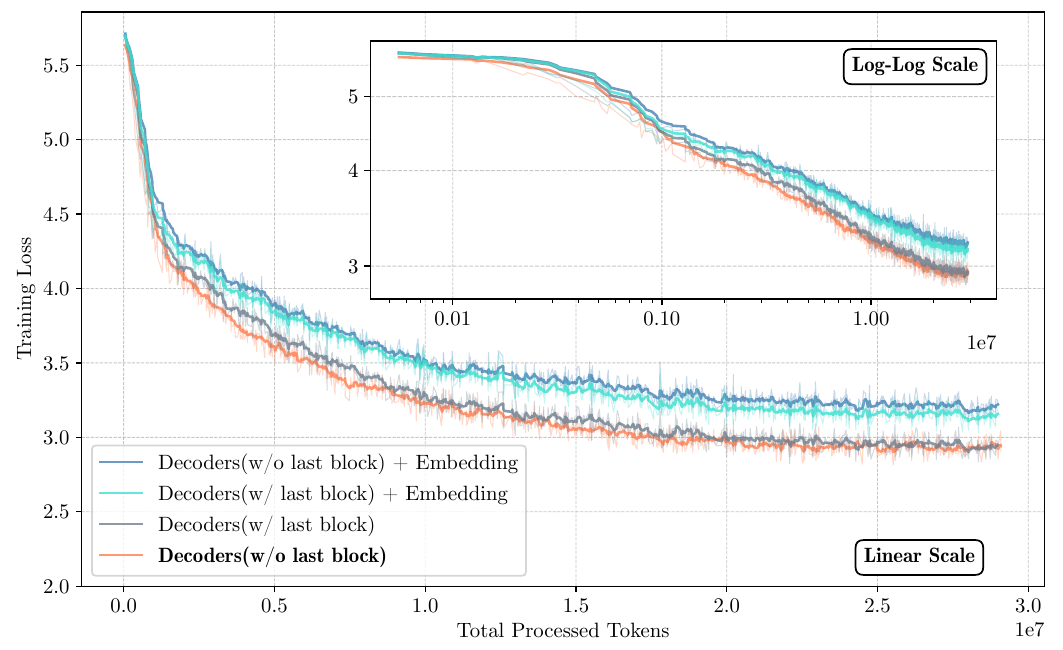}
  \caption{
  \textbf{Performance of the 1B-parameter Transformer with various ODFA scenarios.}
  Training loss curves over total processed tokens for the four ODFA scenarios from experiments. \textit{Decoders(w/o last block)} refers to allocating optical feedback to all decoder blocks except the last one, which uses the ready-calculated gradient from the projected layer directly. \textit{Embedding} indicates that the embedding layer receives the optical feedback, rather than relying on the gradient propagating from the first decoder block, which is also based on optical feedback.
  \textbf{Main axis}, Fully ODFA-enabled training (\textit{Decoder(w/ last block) + Embedding}) shows a modest loss gap but retains stable convergence, confirming ODFA’s contribution. The best performance is obtained with the ODFA scenario adopted in the main text Fig.~\ref{fig:llm}(c). It's also illustrated ODFA has the flexibility to train different Transformer modules on demand.
  \textbf{Inset}, The same data plotted on a log-log scale.
}
  \label{fig:components}
\end{figure*}
To train a GPT-like Transformer using ODFA, we applied several strategies suggested by a previous study~\cite{filipovich2022scaling}.
The general illustration is shown in the main text Fig.~\ref{fig:llm}(a).
To enhance the performance, our ODFA implementation for Transformers deviates from the conventional DFA.
We initially extract the error vector directly from the front of the final projector layer, which is trained using backpropagation. 
Then, the error vector is projected through our optical processor and delivered to the end of each decoder block, excluding the last decoder block and the embedding layer.
The last decoder block receives the exact gradient from the projector layer and propagates the gradient internally using backpropagation.
For the other decoder blocks, we exclusively apply ODFA on a per-block basis.
The end of a decoder block will receive the optical signal as the pseudo error vector for the parameters of the last layer within that block, and then backpropagate the optical signal to the front of the block.
Also, as suggested for DFA in the previous work~\cite{filipovich2022scaling}, we employed activations and residuals asymmetrically.
In the fact that previous work identified that the best activation functions for DFA are continuous and bounded~\cite{launay2019principled}, we switched to the derivative of $\mathrm{tanh}$ activation functions during the backward pass to bridge the gap between ODFA and BP while retaining the classic ReLU activation functions used in Transformers for the forward pass
For the residuals, although remain common in the forward pass, they are disregarded during the backward pass with optical signals to improve alignment between the optical signals and the exact gradients.
Beyond the decoder part, the embedding layer, comprising a token embedding and a position embedding, is trained based on the gradient propagated from the beginning of the first decoder block.
These strategies preserve ODFA's ability to update blocks independently while significantly enhancing performance.

To further validate the precedent approach, in Fig.~\ref{fig:components}, we present additional training results with four ODFA scenarios for the same 1B-parameter Transformer described in the main text Fig.~\ref{fig:llm}. In the first scenario (\textit{Decoders(w/o last block) + Embedding}), we send the optical feedback to decoder blocks (exclude the last decoder block) and also the embedding layer. In the second scenario (\textit{Decoders(w/ last block) + Embedding}), all the decoder blocks and the embedding layer receive the optical feedback, even though a precomputed gradient already exists for the last decoder block. In the third scenario (\textit{Decoders(w/ last block)}), optical feedback is sent to decoder blocks (exclude the last one). And in the fourth scenario (\textit{Decoders(w/o last block)}), we replicate the ODFA scenario used in the main text. For the four scenarios, numbers of parameters directly receiving optical feedback are $331$M, $339$M, $336$M, and $328$M, respectively. From the figure, extended ODFA-enabled training for the language Transformer shows a modest loss gap to our main text ODFA scenario yet converges stably. The best performance is obtained with our main text ODFA scenario. We justify excluding the final projector layer from ODFA because its gradient expansion is essential for the models' rapid alignment with optical feedback. In the fourth scenario (\textit{Decoders(w/o last block)}), training the final projector layer with ODFA leads to an ending training loss of $4.0$, slightly higher than our current performance. Nonetheless, this layer contains only $\sim\!1\%$ of the total parameters. We therefore believe that this trade-off between performance and fully-ODFA training is acceptable. Furthermore, even when the embedding layer is not directly receiving the ODFA signal, it is updated through the gradient flowing from the first decoder block, which is itself trained via ODFA. Thus, the embedding layer is not trained purely by an exact BP gradient propagating from the final layer. Overall, Figure~\ref{fig:components} justifies our chosen ODFA scenario, and more importantly, illustrates ODFA's flexibility to train different Transformer modules on demand.

In addition to the training algorithm, the preferred training configuration also varies between ODFA and BP.
Throughout the main text and this supplementary material, we consistently employed a standardized strategy to determine training configurations
For \textit{SHLW}, \textit{ODFA}, \textit{DFA}, \textit{BP} in the main text Fig.~\ref{fig:llm}(c) and \textit{TDFA} here, we trained the Transformers using ODFA-adopted training configurations.
We first launch the training using ODFA for the aforementioned architecture for $30$ minutes with different training configurations.
The configuration that yielded the best performance at the end of this $30$-minute period was selected as the common configuration for the five aforementioned methods.
In contrast, for \textit{BP}$^*$ in main text Fig.~\ref{fig:llm}(c), we completed training with BP using different training configurations and selected the optimal one as its training configuration.
For the ODFA-adopted configuration, a batch size of $128$ was utilized, along with an Adam optimizer with an initial learning rate set at $0.001$, followed by a cosine learning rate decay. 
A critical distinction between our configuration and the conventional configuration of a language Transformer pertains to the context size. 
We established a context size of $24$ tokens , whereas the typical context size ranges from $500$ to millions ($512$ for BERT, $1024$ for GPT-2, $2048$ for GPT-3, and $128,000$ for Gemini by default). 
For our model with the context window of $24$ tokens, it only looked at the previous $24$ tokens to determine the most likely next token when generating a new token.
Consequently, for a Transformer equipped with a character tokenizer, the likelihood of the next generated token is solely dependent on the preceding $24$ characters or fewer, given the presence of spaces or punctuation. 
This naturally results in a weak linkage between or even within the generated words.
The selection of the context size undeniably has a substantial influence on the performance and capabilities of the language Transformer. 
However, this choice is dictated by the constraints of the model.
Assuming a corpus is tokenized into $N$ tokens with an embedding dimension of $E$ and a Transformer with a context size of $C$ is trained on this corpus, then for one epoch of training, it necessitates $N\times C$ optical projections with a transmission matrix $\underline{\underline{T}}\in \mathbb{C}^{E\times E}$.
Even though our OPU can operate at a very high speed, we still have no benefit from this large number of projections due to the quite small dimension of the projection matrix.

\subsection{Performance of the language Transformers trained by different methods} \label{sec:llmresult}
\begin{figure*}[!htp]
  \centering
  \includegraphics[width=1.\linewidth]{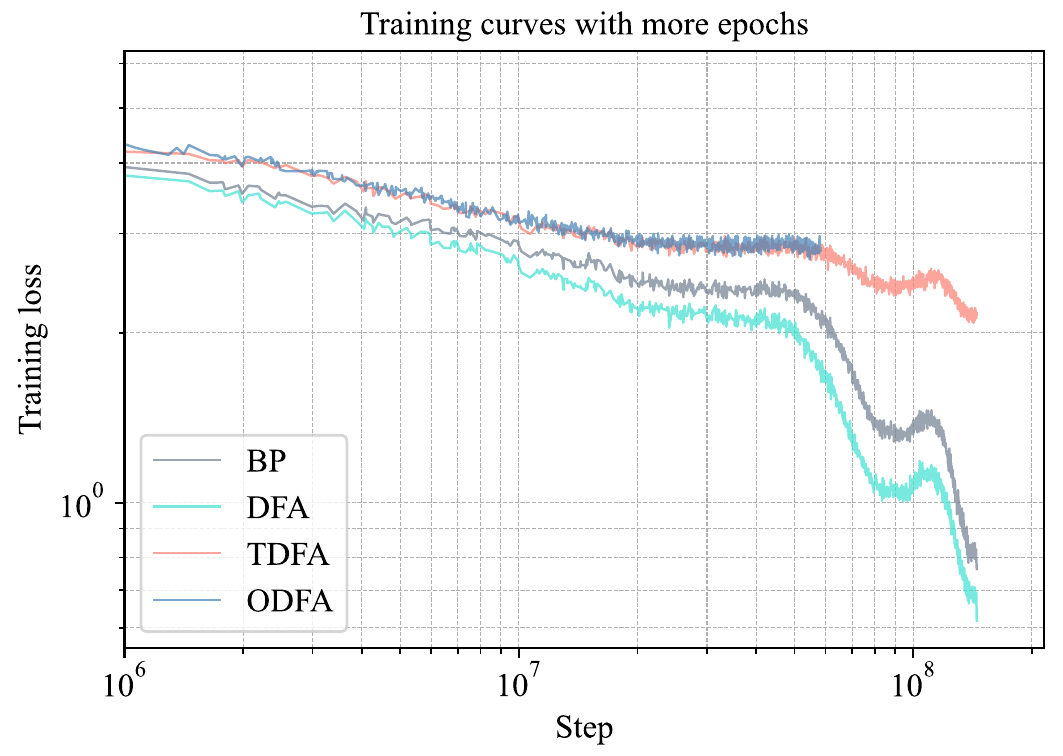}
  \caption{\textbf{Training curves with more epochs}
  The Transformer was trained on a dataset using the BERT-subword tokenizer for five epochs. The Transformer in the figure has the same configuration as the one in the main text Fig.~\ref{fig:llm}. Both axes in the figure are shown on a logarithmic scale.}
  \label{fig:moreepochs}
\end{figure*}

Given the design and structure of the Transformer, we can assess the performances and training durations of the identical Transformer under various training methods to gauge the effectiveness of our ODFA. 
This comparison is crucial in understanding the efficiency and effectiveness of different training methods when applied to the same Transformer architecture.
The training loss curves depicted in Fig.~\ref{fig:tokenize}(a) and Fig.~\ref{fig:moreepochs} represent the models trained by the four methods over two and five epochs, respectively. 
These figures provide a visual representation of the training loss over time, allowing us to observe and analyze the performance of each method.
We noticed that the loss curve of the ODFA training method closely mirrored its digital counterpart, with a minor deviation from the continuous training algorithms as previously observed. 
This observation suggests that ODFA and its digital twin version have similar performance characteristics, which is an important finding in our study.
This deviation was slightly amplified with an increase with the more epochs case. 
This amplification indicates that the performance gap between ODFA and the continuous training algorithms widens with more training epochs. 
In this instance, we didn't implement the optical training of the Transformer for extended epochs, primarily due to the practical constraints of the training duration. 
This decision was made considering the significant time requirement of the optical training method. 
Each epoch of the optical training demanded $10$ hours, while it required approximately half an hour on a high-end GPU ($28$ minutes on Nvidia RTX A6000, $45$ minutes on Nvidia Tesla V100).
Given that TDFA aligns well with ODFA, the additional training of ODFA can be partially inferred by TDFA to a certain degree.
This inference allows us to predict the performance of ODFA for a larger number of epochs without actually performing the training, which can save significant time and resources.

\begin{figure*}[!htp]
  \centering
  \includegraphics[width=1.\linewidth]{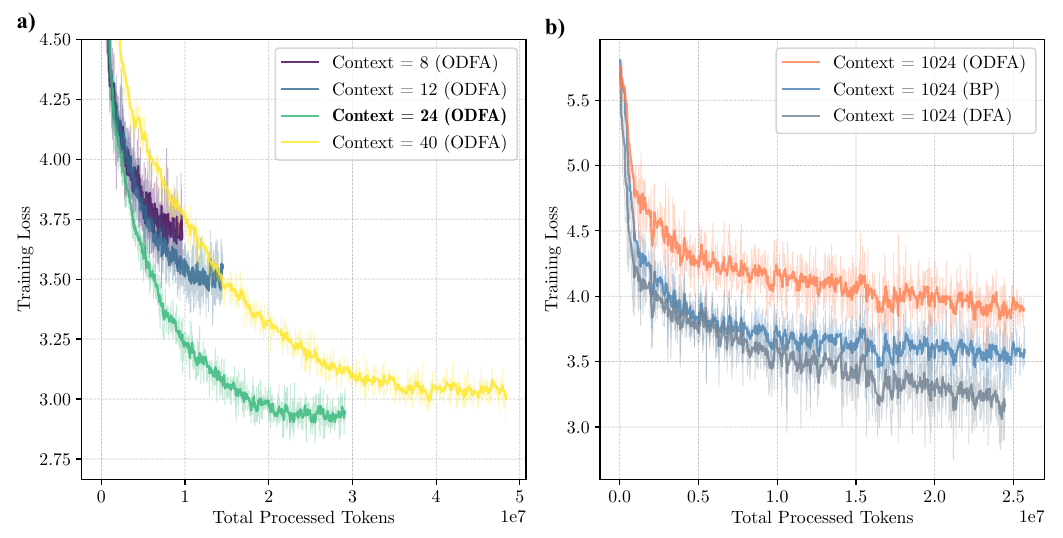}
  \caption{
  \textbf{ODFA training with varying context sizes.} Experimental training using the 1B-parameter Transformer as in the main text Fig.~\ref{fig:llm}(c), but under different context-length configurations.
  \textbf{a}, Training loss curves for context sizes of $8$, $12$, $24$, and $40$ tokens, each run for two epochs. Smaller contexts degrade early performance, while larger contexts do not necessarily yield immediate gains at this stage, aligning with prior observations that extended training is typically needed to realize the benefits of bigger contexts~\cite{kaplan2020scaling}.
  \textbf{b}, Training curves for a context size of $1024$ tokens (equivalent to GPT-2) over short runs ($2$ hours for BP/DFA and $20$ hours for ODFA). All methods follow similar convergence trends and manners as in the main text Fig.~\ref{fig:llm}(c), indicating that ODFA remains compatible with standard context lengths for the low dimensional space in language tasks, subject to ongoing improvements in optical hardware speed.
}
  \label{fig:contextsize}
\end{figure*}

We would like to elaborate further on the extended training duration of ODFA and our selection of a large embedding dimension coupled with a moderate context window. 
The reason for the extended optical training duration is intrinsically linked to our choice of a moderate context window, namely, the number of optical projections.
Each optical projection necessitates a nearly constant exposure duration to capture the optical signal that forms the ODFA signal, irrespective of the projection dimension (optical parallelization). 
This requirement imposes a significant time constraint on the optical training method, which is why we chose a moderate context window and dataset size.

\textbf{Varying context sizes.} Here, in order to provide a more comprehensive assessment of ODFA’s performance when varying context lengths, we evaluated multiple context sizes beyond the original $24$ tokens used in the main text Fig.~\ref{fig:llm}. Our testing involved two types: training for two epochs with smaller context lengths, and running a limited number of steps at a GPT-2-level context size. For the first test (Fig.~\ref{fig:contextsize}(a)), we trained the same Transformer with context lengths of $8$, $12$, $24$, and $40$ using ODFA. Because these experiments occurred at an early training stage, we likewise restrict our conclusions to this phase. Notably, smaller contexts ($8$/$12$ tokens) undermined early performance, presumably because they cannot even capture word-level relationships. Conversely, when increasing the context to $40$ tokens (roughly a sentence in subword-tokens), the model did not immediately improve. We state that the $24$-token context still focuses on word-level interactions, whereas extending to $40$ tokens begins to capture sentence-level structure. Such longer context sequences often converge more slowly but can deliver benefits after extended training, in line with findings from machine learning research~\cite{kaplan2020scaling}. For the second test, we adopted a context size of $1024$ (GPT-2's context size) using BP, DFA, and ODFA. Since the number of tokens processed per epoch scales linearly with context length, training to the same epoch count at $1024$ tokens would exceed one month for ODFA ($100$ hours for BP). Consequently, we trained for a limited number of steps ($2$ hours for BP/DFA and $20$ hours for ODFA), corresponding to approximately $2\%$ of the training in Fig.~\ref{fig:contextsize}(a). Even at this very early stage, all three methods converged at a pace similar to the curves in Fig.~2(c), indicating that ODFA remains stable and compatible with standard Transformer context sizes. Moreover, because language tasks operate at relatively low dimensionality, ODFA’s speed gains rely on continuing improvements in optical hardware. We are actively exploring improved optical hardware and training protocols to further accelerate ODFA, thereby supporting larger context sizes in practical deployments.

\begin{figure*}[!htp]
  \centering
  \includegraphics[width=1.\linewidth]{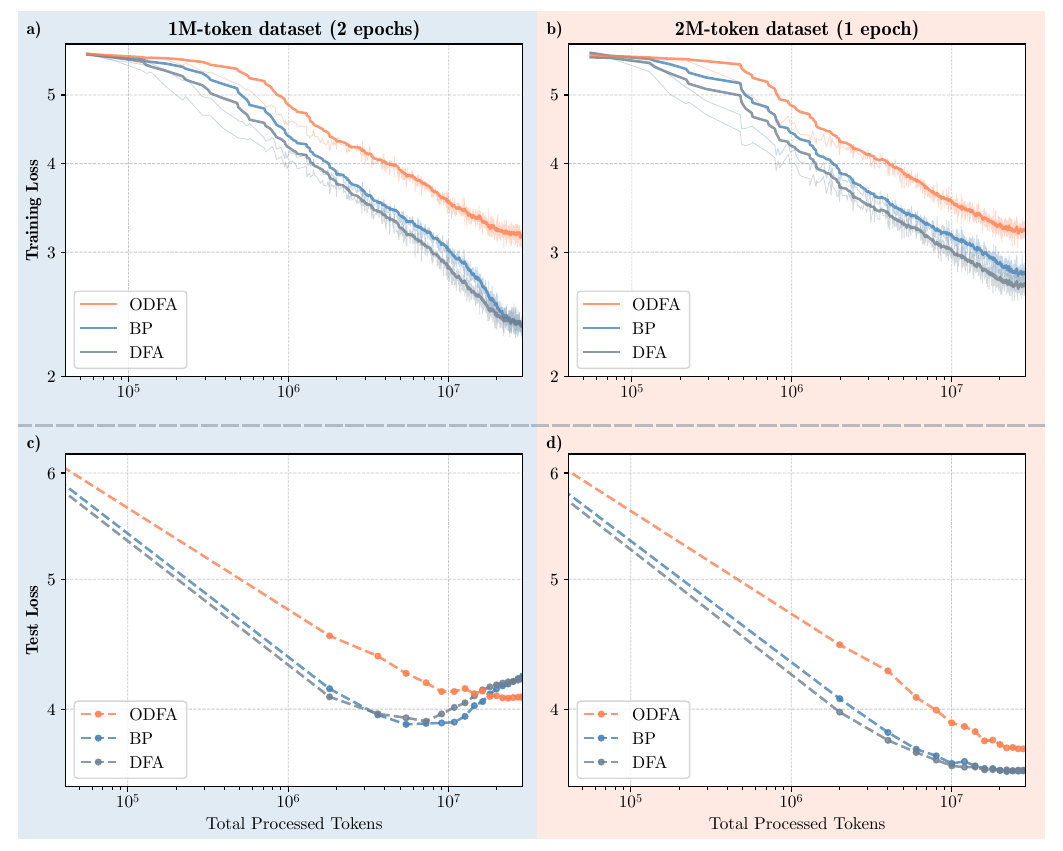}
  \caption{
  \textbf{Training curves and test loss on 1M- vs. 2M-Token Datasets.} A 1B-parameter Transformer is trained using three methods, ODFA (orange), BP (blue), and DFA (gray), on datasets of different sizes. \textbf{a, c} trains on the 1M-token dataset over two epochs, while \textbf{b, d} on the 2M-token dataset over one epoch, resulting in approximately the same total processed tokens. The 2M-token dataset contains all the tokens in the 1M-token dataset. \textbf{a, b} shows the training loss, and \textbf{c, d} gives the test loss on a 122K-token test dataset. All axes are plotted on a log-log scale. \textbf{a}, BP and DFA curves on the 1M-token dataset drop quickly, but eventually show a higher risk of overfitting as in \textbf{c}. This matches the curve in Fig. 16(b) from~\cite{kaplan2020scaling} with a similar training setting (huge model size, limited dataset size, and early training stage). Whereas, ODFA converges more slowly yet avoids severe overfitting at this stage, presenting the lowest test loss. \textbf{b}, on the 2M-token dataset, the final training loss is higher, reflecting fewer passes through the data, yet overall performance remains robust for ODFA. \textbf{d}, all three methods on the 2M-token dataset achieve a notably lower test loss, illustrating improved generalization.
}
  \label{fig:testloss}
\end{figure*}

\textbf{Various dataset volumes.} One frequently asked question concerns the influence of dataset size on training efficacy. As stated in~\cite{kaplan2020scaling}, varying the dataset volume can affect not only training but also test curves throughout the learning process. Hence, we conducted an additional investigation, training the same language Transformer on two differently sized datasets using ODFA/BP/DFA. The first dataset contains 1M tokens, which we employ in the main text Fig.~\ref{fig:llm} and in most investigations in this section (\ref{sec:llm}). Each method trains the Transformer for two epochs on this 1M-token dataset. We then constructed a second dataset by adding another 1M different tokens to the original, forming a 2M-token set, and training for only one epoch. Consequently, runs on the both datasets process roughly the same total number of tokens for a direct comparison of training and test losses. A separate test dataset of 122k tokens, excluded from both the 1M-token and 2M-token sets, was used to evaluate generalization. The results are shown in Fig.~\ref{fig:testloss}.

From the first row, depicting the training losses of ODFA/BP/DFA on both datasets, we observe that larger datasets naturally produce higher early-stage loss. Nonetheless, DFA and ODFA again converge at the same pace as BP, demonstrating no instability for ODFA when confronting a larger dataset. As for the test loss in the second row, an interesting pattern emerges. In Fig.~\ref{fig:testloss}(c), once approximately $6\times 10^6$ tokens are processed, the test loss of BP and DFA begins to rise, indicating overfitting, whereas ODFA, converging more slowly, yet avoids severe overfitting at this point. This apparent divergence echoes prior findings under similar conditions (large model size, limited dataset, early training) in the well-known reference~\cite{kaplan2020scaling}, where the leftmost curves of Fig.~16(b) of the reference also reflect quick overfitting through a growing gap between training and test losses. Furthermore, simply enlarging the dataset can mitigate this overfitting. This outcome is again confirmed for ODFA in Fig.~\ref{fig:testloss}(d), where training with the 2M-token dataset produces a higher final training loss but a lower test loss across BP/DFA/ODFA, supporting that as the case for BP, greater data volume enhances ODFA's generalization.

\textbf{Remarks on ODFA for language Transformers.} A wide context window, large dataset and small embedding dimension, commonly employed in LLM, did not only fall outside the preferred region for our optical training, but it may even reside in a region that is particularly unfavorable or detrimental for our approach.
This observation led us to choose a large embedding dimension and a moderate context window for our study.
Based on this preliminary assessment, we set the embedding dimension of the Transformer to $2040$ to offset the narrow context window and the abbreviated training process. 
This decision was made to compensate for the limitations and to improve the performance.
However, despite this compensation, we must concede that the optical training underperformed BP and the result was still a considerable distance from the benchmark performance. 
This underperformance indicates that there is still room for improvement in the optical training method.
This also catalyzes us to push the optical training to its optimal region (on the climate projection task) to ascertain whether the result can be improved.

\subsection{Impact of noise in the optical training (Large-scale Transformer)} \label{sec:llmnoise}
\begin{figure*}[!htp]
  \centering
  \includegraphics[width=1.\linewidth]{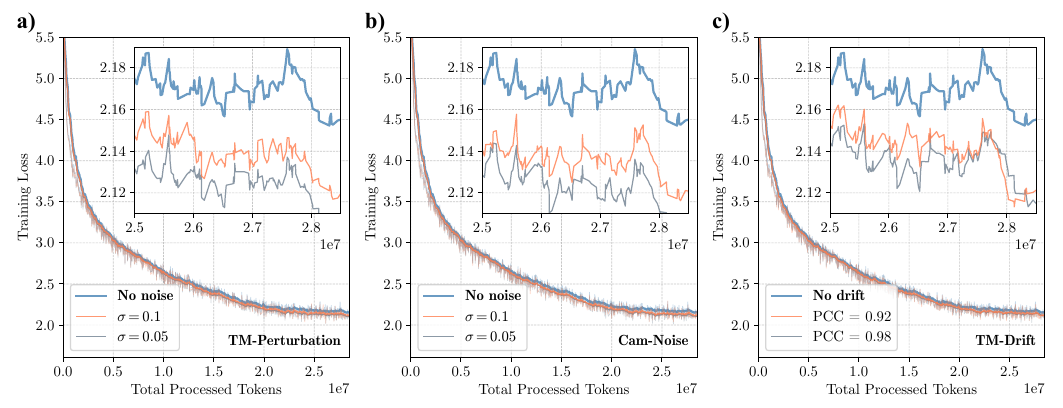}
  \caption{
  \textbf{Noise impact on the 1B-parameter Transformer in the optical training.} Simulations evaluate the impact of three noise types (transmission matrix perturbation, camera measurement noise, and transmission matrix drift), at various magnitudes. All panels show training loss of the same Transformer setting as in the main text Fig.~\ref{fig:llm}(c), with insets highlighting the last few training steps. Identical random seeds across all curves lead to closely matched fine-scale fluctuations. \textbf{a}, TM-Perturbation and \textbf{b}, Cam-Noise confirm that moderate noise levels neither destabilize training as in Fig.~\ref{fig:noise} and Fig.~\ref{fig:dfa2noise}, nor become excessively amplified in large-scale Transformers, aligning with the known regularization benefits of stochastic gradient perturbations~\cite{bottou2012stochastic}. \textbf{c}, TM-Drift similarly illustrates the same potential effect of the noise, with a similar stability to the experiment (PCC = $0.98$), and a more noisy setting (PCC = $0.92$). Although a higher noise level introduces mild degradation, it does not cause catastrophic failure.
}
  \label{fig:llmnoise}
\end{figure*}

Although we initially conducted a basic noise-impact analysis of ODFA on a simple FCNN and the MNIST dataset (SI Note~\ref{sec:expnoise}), uncertainty remained over whether optical noise might be amplified in much larger, more complex models, particularly the 1B-parameter Transformer examined here. We replicated our earlier noise analyses on the full-scale Transformer architecture (Fig.~\ref{fig:llmnoise}). In this test, we introduced the same three types of noise into both the simulated transmission matrix and the random projection during DFA-based training, as what we've done in Fig.~\ref{fig:dfa2noise}: 1. transmission matrix perturbation (TM-Perturbation, $\underline{\underline{T}}(t)=\underline{\underline{T}}(0)+\underline{\underline{\Sigma}}(t)$), 2. camera noise (Cam-Noise, $\underline{s}(t)=\underline{\underline{T}}(0)\underline{e}+\underline{\Sigma}(t)$), 3. transmission matrix drift (TM-Drift, $\underline{\underline{T}}(t)=\underline{\underline{T}}(t-1)+\underline{\underline{\Sigma}}(t)$). For simplicity, we tested three levels (none, small, and moderate) for each noise category, applying identical random seeds so that the fine-scale fluctuations observed in the training curves would align across experiments. These settings mirror the methodology used in the FCNN study, allowing us to more directly compare noise effects across vastly different scales of model complexity.

In the TM-Perturbation and Cam-Noise scenarios, we specifically chose $\sigma=0.05$ (small) and $\sigma=0.1$ (moderate), resulting in SNRs of approximately $400$ and $100$, respectively. As found in Fig.~\ref{fig:noise}, these noise magnitudes led to training outcomes that were slightly better than the no-noise baseline (Fig.~\ref{fig:llmnoise}(a), (b)). Such improvement is consistent with the notion of gradient perturbation~\cite{bottou2012stochastic}, where moderate noise can aid convergence by acting as a regularization mechanism. Indeed, $\sigma=0.05$ outperformed $\sigma=0.05$, indicating that while some level of randomness may be beneficial, excessive noise begins to hinder optimization. We further evaluated TM-Drift, using $\text{PCC}=0.98$ (same level with our experiments, see SI Note~\ref{sec:expstab}), and $\text{PCC}=0.92$, in Fig.~\ref{fig:llmnoise}(c). Although this drift-based noise produced lower gains than the other two noise types, it still offered a modest advantage over a no-noise setup. Thus, these results reinforce that neither small nor moderate noise is harmful at the billion-parameter scale; in certain cases, it can even enhance early convergence.

Overall, our simulations suggest that moderate levels of optical noise do not severely affect performance in large-scale Transformers. Rather, small random noise sometimes helps initial training and does not destabilize the further learning process. Although very high noise can hurt performance, we did not observe catastrophic failure even with a higher noise level than our optical experiment. Consequently, the noise levels typical of our experimental conditions appear compatible with stable and effective learning in billion-parameter language Transformers. We also emphasize that we did not fine-tune the noise parameters to optimize performance; we simply adopted the same parameter settings used in the MNIST study.

\section{ODFA for vision tasks} \label{sec:vit}
\subsection{Preprocessing for Climate projection task and configurations} \label{sec:vitdata}
\begin{figure*}[!htp]
  \centering
  \includegraphics[width=1.\linewidth]{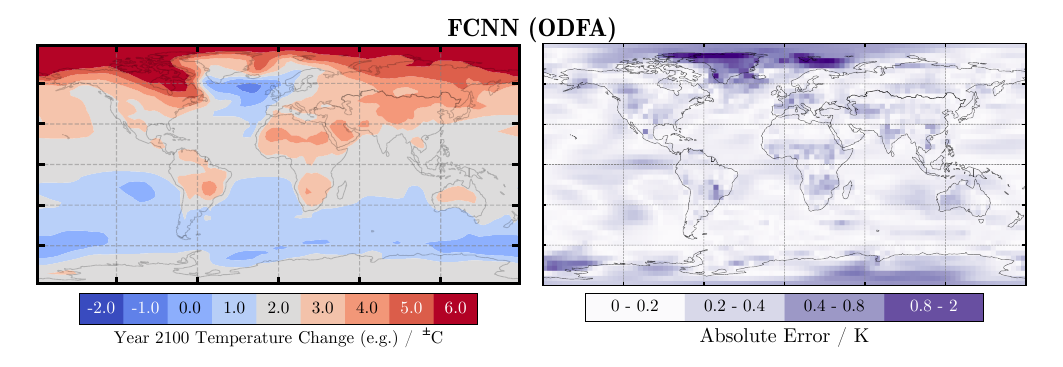}
  \caption{
  \textbf{FCNN trained by ODFA on Climate projection task.} A 1.3B-parameter FCNN trained by ODFA. The prediction of the global temperature change for Year 2100, and the absolute error relative to the ground truth are with the same color scales used in the main text Fig.~\ref{fig:climate}(b) and Fig.~\ref{fig:climate}(d). The similar low residual differences reinforce ODFA's effectiveness in training extensive FCNN.
}
  \label{fig:vitfcnn}
\end{figure*}

As discussed previously, the most fitting tasks for our ODFA implementation are those with a small number of samples and a large number of dimensions, typically seen in domains such as astronomy and geometry.
Here, we examined the applicability of a hardware-software co-design approach in the context of climate projection, using the ClimateBench dataset as our experimental platform. 
This dataset, tailored for climate modeling, comprises a modest number of samples, totaling $894$ instances in the training set. Each sample is characterized by a high-dimensional input space of $55$k dimensions and an output space of $14$k dimensions, making it well-suited for investigating the efficacy of our codesigned hardware-software architecture.
We explored two architectures: Vision Transformer (ViT) and FCNN. 
The ViT configuration involved downsampling the input data to $5$k and encoding to an embedding dimension of $2048$, employing $8$ decoder blocks, a history length (context size) of $10$, $16$ attention heads per decoder block, and an MLP layer structure of $[2048, 4096, 2048]$ within each decoder block. 

Training of the ViT architecture using ODFA is very similar to the language Transformer.
In the ViT context, ODFA was again applied exclusively to the decoder part, where the embedding layers receive optical signals from the projector layer, and subsequently propagated through the embedding layers using backpropagation.
For the ViT architecture we used~\cite{Nguyen2023} for the climate projection task, the decoder part is similar to our previous implementation for the language task.
The embedding layers, however, incorporated a variable-separate tokenizer, variable aggregation, and time-position embedding layer.
Given that the primary difference between ViT here and GPT-like Transformers resides in the embedding layers, which has minimal influence on the decoder part, we adopted the same algorithm and strategy for training the ViT as employed for the language model.
Training of the ViT architecture utilized a learning rate of $0.001$ with the Adam optimizer, spanning $2$ epochs. 
Training times for the architectures amounted to $53$ minutes for BP, DFA, and TDFA, while ODFA required approximately $13$ hours.
Conversely, the FCNN architecture was trained directly on the raw input and output data, employing a layer configuration of $[55,296, 13,824, 13,824, 6,912, 13,824]$, with the tanh activation function utilized across layers.
As a complement to the main text Fig.~\ref{fig:climate}, we here present the ODFA-trained FCNN’s predictions and corresponding absolute errors in Fig.~\ref{fig:vitfcnn}, employing the same colorbar from Fig.~3 for consistency. Despite the substantial model size and the direct mapping from high-dimensional climate data, ODFA successfully learns to predict the year 2100 temperature change with a close match to ground truth. Although the quantitative metric indicates that FCNN(ODFA) is less accurate than ViTs, the outcome still further underscores ODFA’s versatility and scalability, particularly in scenarios where fully connected architectures can capture rich global dependencies.

\subsection{Evaluations of the ViTs trained by different methods} \label{sec:vitmetric}
\begin{figure*}[!htp]
  \centering
  \includegraphics[width=1.\linewidth]{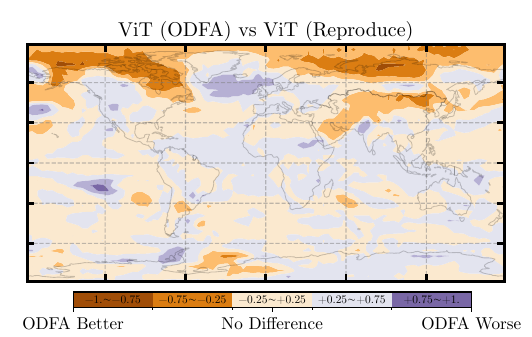}
  \caption{
  \textbf{Grid-by-grid comparison of ViT trained by different methods.} A global difference map contrasts ViT(ODFA) and ViT(Reproduce) from the data in the main text Fig.~\ref{fig:climate}. Each grid is colored based on which model has a smaller error relative to the ground truth: orange denotes ODFA performs better, purple indicates worse, and Papaya Whip for nearly no difference.
}
  \label{fig:vitbitbetter}
\end{figure*}

Although we presented the predictions and errors of ViT(ODFA) and ViT(Reproduce) in the main text Fig.~\ref{fig:climate}, the difference between their errors may still not be fully apparent. Therefore, to improve clarity, we visualize their performance difference in another way. In Fig.~\ref{fig:vitbitbetter}, we compare the absolute errors of ViT(ODFA) and ViT(Reproduce), focusing on which model's prediction is closer to the target. While this approach does not yield a definitive conclusion, it suggests a qualitative impression that ViT(ODFA) and ViT(Reproduce) perform comparably overall, differing mainly in regional temperature predictions. Neither model shows extreme biases, and any error variations remain within a reasonable range.

\begin{figure*}[!htp]
  \centering
  \includegraphics[width=1.\linewidth]{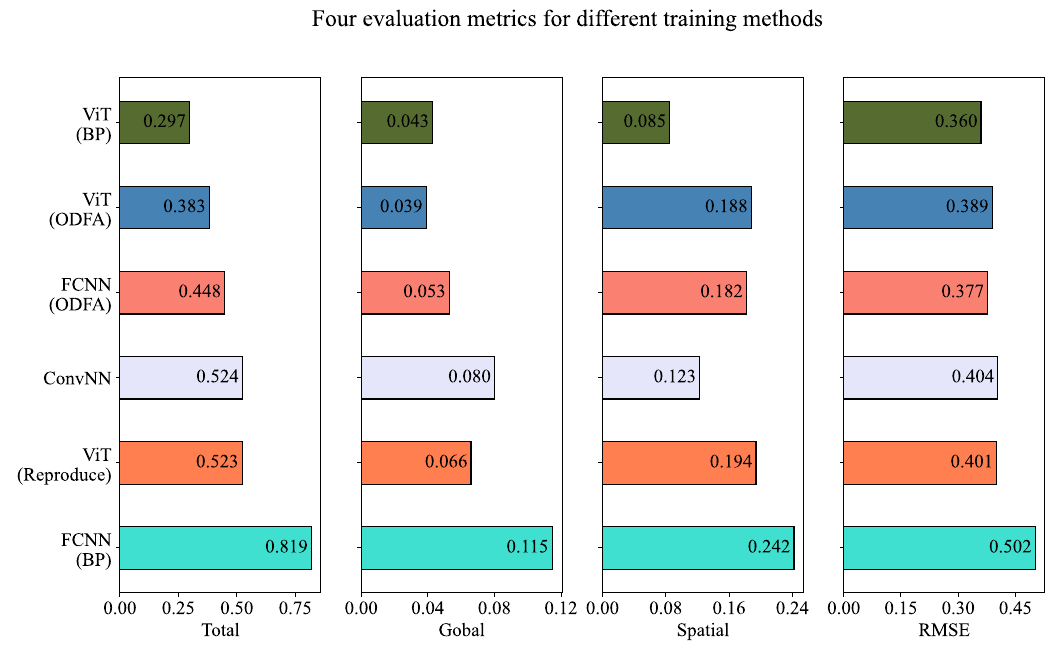}
  \caption{\textbf{Performances of BP- or ODFA-trained models over four RMSE-based metrics.}
  Besides ODFA, the BP results of \textit{ViT(Reproduce)} were obtained using the same configuration of \textit{ViT(ODFA)}, and the results of \textit{ConvNN} are from~\cite{Watson-Parris2022}.
  We also introduced a Shallow Training (\textit{SHLW}) baseline for the ViT here with the last decoder block trained by backpropagation and all the others frozen. To maintain figure clarity and readability, the \textit{SHLW} is presented in this caption. For \textit{SHLW}, the four evaluation matrix values are $4.511$, $0.733$, $0.848$, and $1.756$, respectively.
  }
  \label{fig:metrics}
\end{figure*}

To quantitatively evaluate model performance, we employed four evaluation metrics: spatial normalized root-mean-square error (Spatial), global mean (Global), weighted total mean (Total), and root-mean-square error (RMSE). 
Spatial NRMSE quantifies the spatial bias between the temporal mean of the prediction $y$ and the target $\tilde{y}$ and is defined as,
\begin{equation}
    \mathrm{Spatial} = \left.\sqrt{\left<\left(\frac{1}{T}\sum_{t=1}^T{y} - \frac{1}{T}\sum_{t=1}^T{\tilde{y}}\right)^2\right>}\middle/\frac{1}{T}\sum_{t=1}^T\left<y\right>\right. ,
\end{equation}
where $\left<y\right>$ is the global mean of $y$ (or $\tilde{y}$):
\begin{equation}
    \left<y\right>=\frac{1}{H\times W}\sum_{i=1}^H\sum_{j=1}^W{\sin{\frac{\pi \times i}{H}}y_{ij}}.
\end{equation}
The global NRMSE to measure the bias between the global mean of $y$ and $\tilde{y}$ is defined as,
\begin{equation}
    \mathrm{Global} = \left.\sqrt{\frac{1}{T}\sum_{t=1}^T\left(\left<y\right> - \left<\tilde{y}\right>\right)^2}\middle/\frac{1}{T}\sum_{t=1}^T\left<y\right>\right. .
\end{equation}
The total RMSE is the weighted sum of the aforementioned two metrics,
\begin{equation}
    \mathrm{Total} = \mathrm{Spatial} + \alpha \cdot \mathrm{Global},
\end{equation}
where $\alpha$ is suggested to be $5$.
The four metrics' results are shown in Fig.~\ref{fig:metrics}.

Here, we would like to explain more on the different methods.
Among the seven methods shown in Fig.~\ref{fig:metrics}, the two benchmarks \textit{ViT(BP)}~\cite{Nguyen2023} and \textit{ConvNN}~\cite{Watson-Parris2022} are directly cited from the existing references.
The remaining five methods, \textit{ViT(ODFA)}, \textit{FCNN(ODFA)}, \textit{ViT(Reproduce)}, \textit{FCNN(BP)}, and \textit{SHLW}, are implemented by ourselves.
Different methods for ViT faced identical architectures.
Notably, although \textit{ViT(BP)} and \textit{ViT(Reproduce)} are both trained using backpropagation, they applied distinct training configurations.
Following a consistent approach with the language task, we categorized the training configuration of \textit{ViT(BP)} and \textit{ConvNN} as BP-adopted configurations, and those for \textit{ViT(ODFA)}, \textit{FCNN(ODFA)}, \textit{ViT(Reproduce)}, \textit{FCNN(BP)}, and \textit{SHLW} as ODFA-adopted configurations.
To select the ODFA-adopted training configurations, we adhered to the same strategy used for the language task: identifying the configuration that exhibited the best performance after 30 minutes of optical training using ODFA.

From the figure, it is evident that the ViT benchmark achieved the highest performance, while \textit{FCNN(BP)} exhibited the lowest, excluding \textit{SHLW}.
While it is not surprising that no ODFA-trained model surpassed the benchmark in terms of performance, the results obtained by ODFA for this high-dimensional task were notably closer to the benchmark compared to the language task.
To isolate the impact of the advanced architecture, we can first examine the FCNN results. 
The extra-large dimension of the output layer appears to have hindered the performance of BP under the ODFA-adopted training configuration.
The big gap between the \textit{FCNN(ODFA)} and \textit{FCNN(BP)} indicates that the paradigms for selecting the training hyperparameters are significantly different for ODFA and BP in the extra-large region.
This observation is further supported by the gap between \textit{ViT(ODFA)} and \textit{ViT(Reproduce)}.
Regarding the ViT architecture, we can observe a similar trend.
Under the ODFA-adopted training configuration, \textit{ViT(ODFA)}  approached the ViT benchmark more closely, likely due to the benefits of the advanced architecture.
It is important to emphasize that ODFA did not outperform BP even in this preferred large-scale task.
The primary conclusions drawn from these observations are: first, ODFA can achieve a good performance close to what BP can have within this preferred large-scale task; and second, the paradigms for determining hyperparameters for ODFA and BP are distinct and require careful consideration before launching the training.

\section{ODFA on diffusion models} \label{sec:dit}
So far, we have shown that ODFA can effectively train both language and vision Transformers by leveraging hybrid optical feedback, demonstrating its adaptability across diverse modalities. Yet, modern Transformer research extends well beyond these domains, particularly into diffusion-based generative methods that iteratively transform noise into structured outputs. A class of diffusion models based on the Transformer architecture, so-called Diffusion Transformer (DiT)~\cite{peebles2023scalable}, inherits excellent scaling properties. Demonstrating ODFA on a latent DiT is therefore a logical extension of our earlier work, as it tests whether ODFA remains effective on a more extensive framework. If ODFA continues to perform well on this complex paradigm, it further validates the use of optical feedback methods for advanced generative tasks.

Throughout this section (\ref{sec:dit}), we reference~\cite{peebles2023scalable} for the original source from which we borrowed both the DiT framework and the adaLN-Zero block. We train a standard DiT-B/2 using ODFA on MNIST and the Animal Face dataset to investigate its flexibility in varying task difficulties.

\subsection{ODFA for Diffusion Transformers (DiT)} \label{sec:ditarch}
\begin{figure*}[!htp]
  \centering
  \includegraphics[width=1.\linewidth]{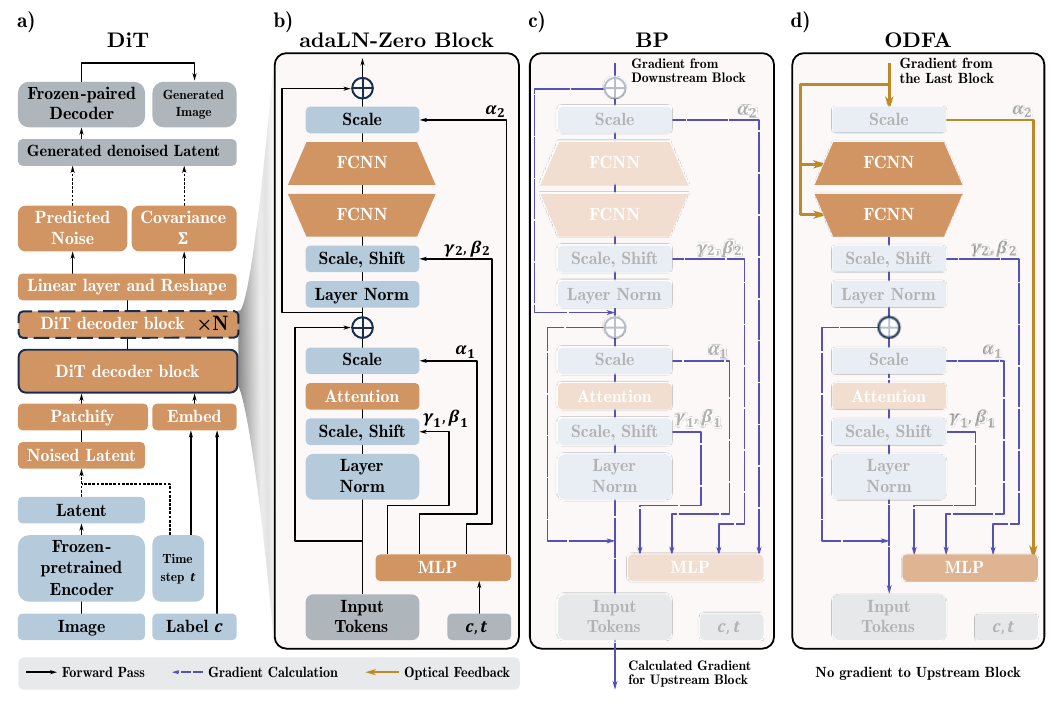}
  \caption{
  \textbf{Diffusion Transformer (DiT) architecture and integration of ODFA.}
  \textbf{a}, Schematic of the latent diffusion model, adapted from~\cite{peebles2023scalable}, where the main DiT module (orange) operates on latent representations derived from a frozen-pretrained encoder-decoder pair (gray). The diffusion (sky blue) and reverse diffusion (gray) processes occur in latent space, so a purely ``random" latent does not appear as uncorrelated pixel-level noise, but probably color lumps. Black arrows indicate the forward pass through the model.
  \textbf{b}, Detailed view of the adaLN-Zero DiT block~\cite{peebles2023scalable}, illustrating the modules with parameters (orange), the near-parameter-free operations (sky blue), and the input data flow (gray). The block incorporates conditioning via adaptive layer norm from MLP.
  \textbf{c}, Standard BP to train the DiT block, in which dashed blue arrows represent the full gradient flow through each module. The gradient must be computed at every trainable module, from downstream to upstream layers. Each dashed blue arrow indicates fully accurate gradient calculations.
  \textbf{d}, ODFA training strategy, highlighting the partial replacement of gradient calculations with optical feedback (orange arrows), whereas modules with partial non-transparency indicate reduced gradient computation workload. No gradients are propagated from downstream blocks in ODFA; instead, the optical feedback is injected directly into each trainable block. This arrangement reduces computational overhead while preserving stable training, extending ODFA’s applicability.
}
  \label{fig:ditarch}
\end{figure*}

DiT models replace the conventional U-Net backbone with a Transformer architecture, outperforming prior models. They employ an iterative generative process that refines an initially noisy latent state until a coherent structure emerges. Unlike pixel-space diffusion methods, which apply forward and reverse diffusion on raw image grids, the latent DiT~\cite{peebles2023scalable} uses a pretrained encoder–decoder pair to extract latent features. 

Figure~\ref{fig:ditarch}(a) shows the core design of the DiT with color-coding to emphasize three main processes: \textbf{1.} The latent diffusion process (sky blue), a forward noising process which gradually applies noise to a latent $z_0=\mathcal{E}(x_0)$, where $\mathcal{E}$ is a frozen encoder to extract features and $x_0$ is the raw real data. The sample is given by $z_t=\sqrt{\bar{\alpha}_t}z_0+\sqrt{1-\bar{\alpha}_t}\epsilon_t$, where $\bar{\alpha}_t$ is the retention factor, and $\epsilon_t\sim\mathcal{N}(0,\mathbf{I})$; \textbf{2.} The core DiT modules (orange) to learn the diffusion process. It aims to give the predicted noise $\epsilon_{\theta}(z_t)$ and covariance $\Sigma_{\theta}(z_t)$; \textbf{3.} The reverse diffusion steps (gray), denoising the latent based on $\epsilon_{\theta}(z_t)$ and $\Sigma_{\theta}(z_t)$, to give a generated data $\hat{x}_0=\mathcal{D}(\hat{z}_0)$, where $\mathcal{D}$ is a pretrained-frozen decoder. Meanwhile, because the DiT works in latent space through the pretrained-frozen encoder-decoder ($\mathcal{E}$-$\mathcal{D}$) pairs, purely noised latents often appear as color patches rather than pixel-level randomness. Within Fig.~\ref{fig:ditarch}(a), the forward pass arrows (in black) track how the latent $z$ is patchified, embedded, and processed through multiple stacked DiT decoder blocks. We adopt the same adaLN-Zero block used in the original paper, conditioned on noise timesteps $t$ and class labels $c$ through a multilayer perceptron (MLP). The internal structure of an adaLN-Zero block appears in Fig.~\ref{fig:ditarch}(b). Black arrows again represent the forward pass. Orange elements mark layers with learnable parameters (attention layers, FCNNs, MLPs). Despite sky-blue transformation modules contain some parameters, they are comparatively minimal. The block's input is shown in gray. This color scheme parallels Fig.~\ref{fig:ditarch}(a), yet maintains a clear distinction between the primary trainable components and those requiring fewer adjustments. 

To integrate ODFA into the training of DiT, we first examine the conventional BP flow depicted in Fig.~\ref{fig:ditarch}(c). Dashed blue arrows denote the gradient flow during the backward pass. Each learnable module receives gradients from downstream layers and propagates them upstream once the gradient computation finishes. As for ODFA (Fig.~\ref{fig:ditarch}(d)), optical feedback (orange arrows) reduces the need for explicit gradient updates from downstream modules. Each DiT decoder block directly receives optical feedback signals from the final linear layer rather than a full gradient, while some connections still rely on standard gradients (dashed blue arrows). In particular, the attention layer demands conventional gradient flow, whereas other parts incur lighter computational loads. The transparency of the main learnable modules reflects this lowered overhead (higher transparency implies smaller reductions). The ODFA integration on DiT architecture enables us to evaluate ODFA’s adaptability for another Transformer variant to test its performance variability.

\subsection{DiT trained with ODFA on MNIST dataset} \label{sec:ditmnist}
\begin{figure*}[!htp]
  \centering
  \includegraphics[width=1.\linewidth]{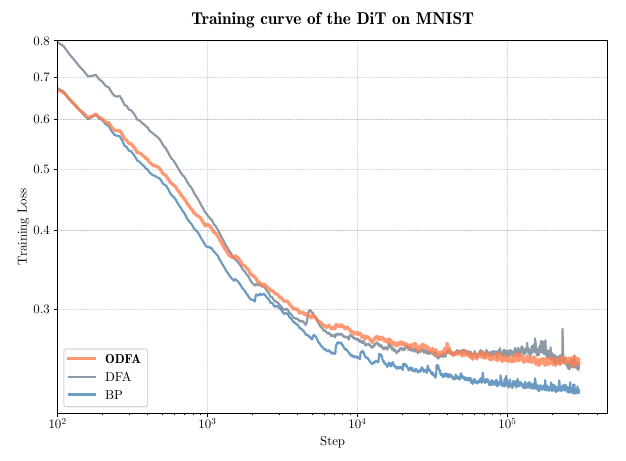}
  \caption{
  \textbf{Training curves of a Diffusion Transformer with BP, DFA, and ODFA.} A DiT-B/2 is trained on MNIST using ODFA, DFA, and BP, respectively. The latent features are extracted by the off-the-shelf pre-trained ft-EMA VAE encoder, given the $60$k digit images. The encoder is frozen during the whole training. The model contains $129$ million parameters, with 35 million directly receiving optical signal in ODFA. All three methods, ODFA (orange), DFA (gray), and BP (blue), run for $300$k steps. ODFA requires $46.9$ hours of training compared to $6.7$ hours for BP and DFA, yet this time ratio is smaller than that observed for the 1B-parameter language Transformer, suggesting that ODFA’s computational overhead can be reduced across model architectures.
  \textbf{Main axis}, Training loss curves in log-log scale. 
  Although BP converges to the lowest final loss, both DFA and ODFA closely track its trajectory, where the performance gap remains modest, indicating that DFA and ODFA effectively train the DiT. 
}
  \label{fig:ditmnloss}
\end{figure*}

In this subsection, we conduct an experimental evaluation of ODFA on a class-conditional DiT trained with the MNIST dataset. Our primary goal is to examine whether ODFA and DFA can approximate BP’s training convergence and generative quality, even in a simpler image domain. Although MNIST is commonly regarded as a less challenging dataset, it still serves as a stepping stone for validating whether the optical feedback can effectively train DiT in latent space. We trained a standard DiT-B/2 model, where the latent has $4$ input channels, the DiT has an embedding size of $768$ with a patch size of $2$, and there are $12$ adaLN-Zero blocks with $12$ heads. For each decoder block, the FCNN has three layers, $[768, 3072, 768]$, and an MLP of $[768, 4608]$. The original gray-scale MNIST image with shape $28\times28$ is resized to an RGB image $32\times32\times3$ before applying the encoder. Following the training setting in~\cite{peebles2023scalable}, we used off-the-shelf, pretrained Variational Autoencoders (VAEs). We freeze the encoder–decoder pair (sd-vae-ft-ema) throughout training, initialize the final linear layer with zeros, and adopt standard weight initializations for all remaining parameters. We train every model with AdamW at a fixed learning rate of $2\times 10^{-4}$, no weight decay and a batch size of $32$ on a single GPU. Due to time constraints, we limit training to $3\times10^5$ steps (where the original paper trains effectively $6\times10^6$ steps) without hyperparameter tuning to have an ODFA-adapted configuration.

Figure~\ref{fig:ditmnloss} provides a comparison of training loss trajectories for the three methods: ODFA (orange curve), DFA (gray curve), and standard BP (blue curve). Although BP converges faster overall, ODFA and DFA both show consistent declines in loss, verifying they can guide the DiT in generating coherent latent updates. ODFA completes training in $46.9$ hours, whereas BP takes $6.7$ hours, an approximately $7\times$ increase in time. However, this ratio is still an improvement over the difference observed in the 1B-parameter language Transformer (\ref{sec:llm}), where ODFA’s time cost was proportionally larger. 

\begin{figure*}[!htp]
  \centering
  \includegraphics[width=1.\linewidth]{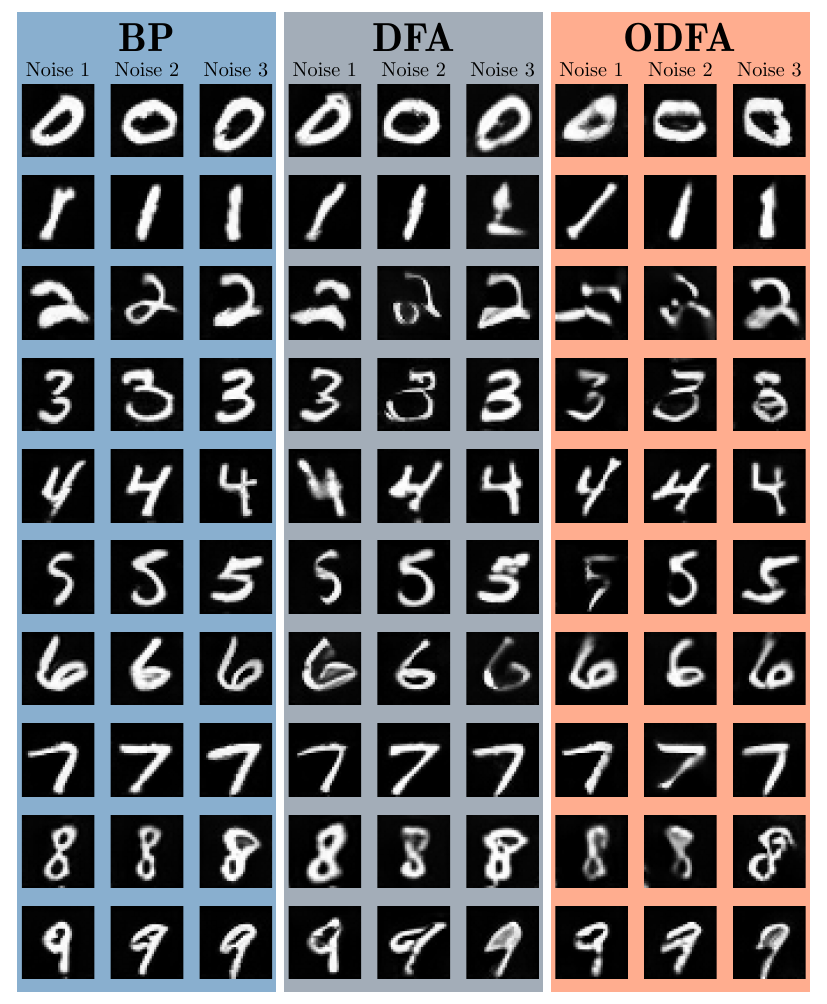}
  \caption{
  \textbf{Class-conditional MNIST samples from DiT-B/2 models trained by different methods.} Each column corresponds to the same initial fully noised latent (details in Fig.~\ref{fig:ditmnchain}).
}
  \label{fig:ditmngen}
\end{figure*}

While the training loss curves offer a quantitative snapshot, generative models ultimately need to produce sharp and semantically accurate outputs. Figure~\ref{fig:ditmngen} shows class-conditional samples from the DiT-B/2 trained with each method, using three different noise seeds (Noise 1, Noise 2, Noise 3) for every digit class. We apply a classifier-free guidance scale of $1.5$ using $180$ DDPM steps. Each column corresponds to an identical initial noisy latent state, and each row to the guided class, allowing a side-by-side visual comparison. In many cases, all three methods generate recognizable digits, indicating that the model can recover fundamental structure from the latent noising process. Closer zoom-in suggests that BP’s outputs exhibit slightly clearer boundaries in some instances, particularly for digits such as "2", "6", or "8". Meanwhile, ODFA/DFA’s digits remain structurally correct but can show small detailed degradation, consistent with the modest gap in training loss. ODFA’s outputs are sometimes than DFA’s at the margins. Overall, however, ODFA remains capable of producing sufficiently clear digits.

\begin{figure*}[!htp]
  \centering
  \includegraphics[width=.95\linewidth]{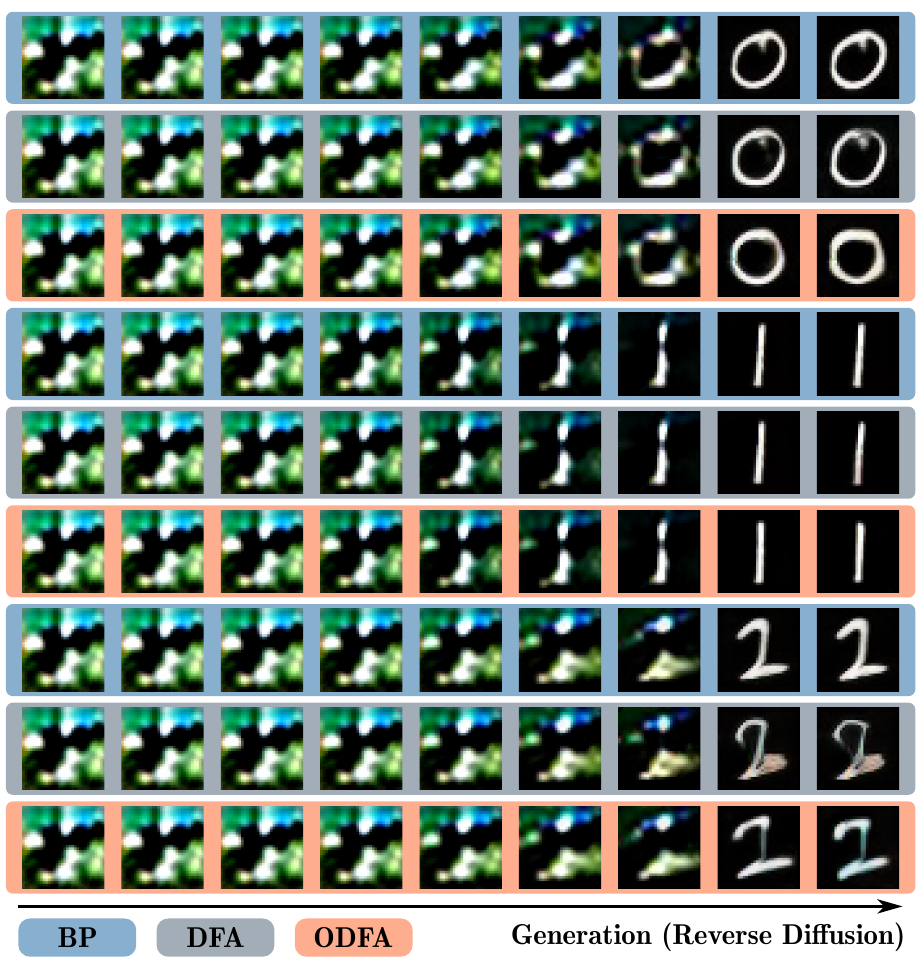}
  \caption{
  \textbf{Reverse diffusion chains of DiT-B/2 trained by different methods on MNIST.}
  Each row presents the sequential evolution of a class-conditioned (three classes, respectively) image from a fully noised starting point (left) to a final digit (right), with the method indicated by the background color. Because the DiT operates in latent space, the image of the noised latent after the pretrained-frozen decoder often appears as blurred color patches. Samples are actually RGB images, as subtle color remnants remain in some ODFA and DFA samples, reflecting the modest loss gap in Fig.~\ref{fig:ditmnloss}. 
  Samples are generated with a classifier-free guidance scale of $1.5$, using $180$ DDPM steps and a ft-EMA VAE decoder.
}
  \label{fig:ditmnchain}
\end{figure*}

To further analyze the generative dynamics, Figure~\ref{fig:ditmnchain} displays reverse diffusion chains for three digit classes, comparing how each method transforms a fully noised latent into a final reconstructed digit. Each row presents snapshots at multiple intervals throughout the diffusion reversal. The leftmost images reveal the heavily noised latent as decoded by the frozen VAE, frequently displaying an array of blurred color patches, reflecting that the real noise is introduced in latent space. At intermediate steps, the generative process gradually converges to features corresponding to the digit outline, until the image on the rightmost side becomes a clear digit (RGB image) that matches the target class. For BP, the transition often appears slightly smoother, while ODFA’s paths can show small fluctuations in local color or shape. However, these differences rarely degrade the final result. The observed consistency in the final steps indicates that ODFA is stable during the reverse diffusion process.

\subsection{DiT trained with ODFA on Animal Faces dataset} \label{sec:ditanimal}
\begin{figure*}[!htp]
  \centering
  \includegraphics[width=1.\linewidth]{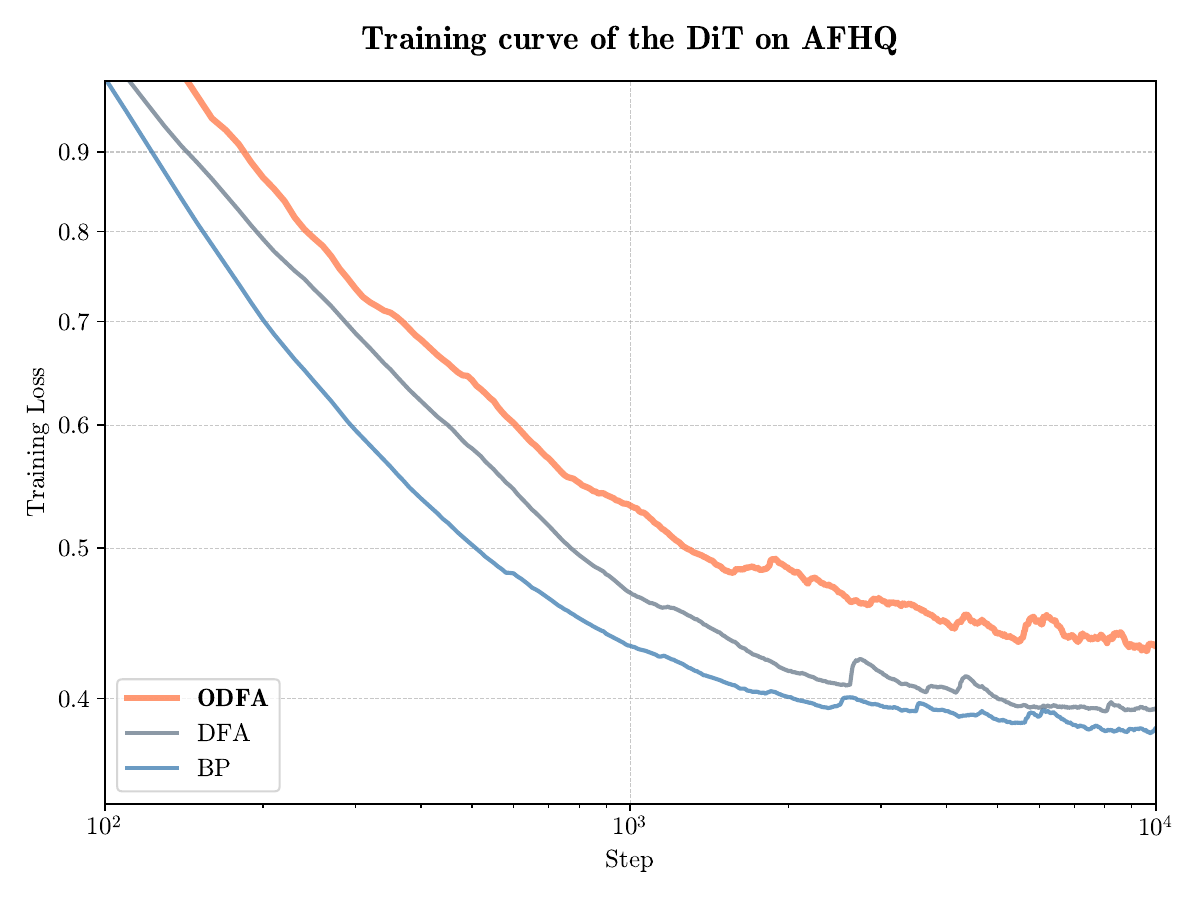}
  \caption{
  \textbf{Training curves of a Diffusion Transformer on Animal Faces dataset.} A DiT-B/2 is trained on AFHQv2-64~\cite{choi2020stargan} using ODFA, DFA, and BP, respectively. AFHQv2 contains $10$k cat/dog face images, whose original resolution is $512\times512$ and then downsampled to $64\times 64$ during the preprocessing. The latent features are extracted by the off-the-shelf pre-trained ft-EMA VAE encoder. The encoder is frozen during the whole training. The model contains $129$ million parameters, with $35$ million directly receiving optical signal in ODFA. All three methods, ODFA (orange), DFA (gray), and BP (blue), run for $10$k steps. ODFA requires $14.2$ hours of training compared to $2.1$ hours for BP and DFA. 
}
  \label{fig:ditafloss}
\end{figure*}

Having demonstrated ODFA’s feasibility on the MNIST dataset, we now extend our investigation to a more diverse, structurally complex domain using the Animal Faces (AFHQv2) dataset~\cite{choi2020stargan}. While MNIST offers a simple benchmark focusing on digits, AFHQv2 includes a diverse collection of cat and dog faces at an original resolution of $512\times512\times3$, subsequently down-sampled to $64\times64\times3$ in this test. This introduces larger color variations, facial structures, and wider style differences, providing a suitable task to examine whether ODFA maintains training stability and generative fidelity when confronted with more sophisticated inputs.

AFHQv2 contains $10$k images covering cats and dogs of different species. As in the MNIST experiments, we employ DiT-B/2, featuring $129$ million parameters. Among these, $35$ million parameters directly receive optical feedback signals under ODFA. As before, a pretrained-frozen encoder-decoder pair (sd-vae-ft-ema) is used to transfer the images into latent space. We repetitively apply an AdamW optimizer at a fixed learning rate of $2\times 10^{-4}$, no weight decay, and a batch size of $64$ here on a single GPU for $10^4$ steps. In terms of training time, ODFA took $14.2$ hours to complete whereas BP finished in $2.1$ hours under the same computer. The training time ratio is $6.7\times$, maintaining the same level as on the MNIST dataset. Figure~\ref{fig:ditafloss} shows training curves for ODFA (orange), DFA (gray), and BP (blue) on AFHQv2. Despite the curves are plotted over a shorter range than the MNIST experiments, they can also give clear trends in loss reduction and learning pace. Again, BP's loss decreases more rapidly, resembling its performance advantage similar in the MNIST setting. DFA tracks BP’s slope, while ODFA’s curve remains visibly higher, reflecting a larger error due to quantization for a more complex task. However, the differences among the three methods stabilize as training proceeds, suggesting that ODFA can effectively refine the model’s weights over time even for such a complicated task. 

\begin{figure*}[!htp]
  \centering
  \includegraphics[width=1.\linewidth]{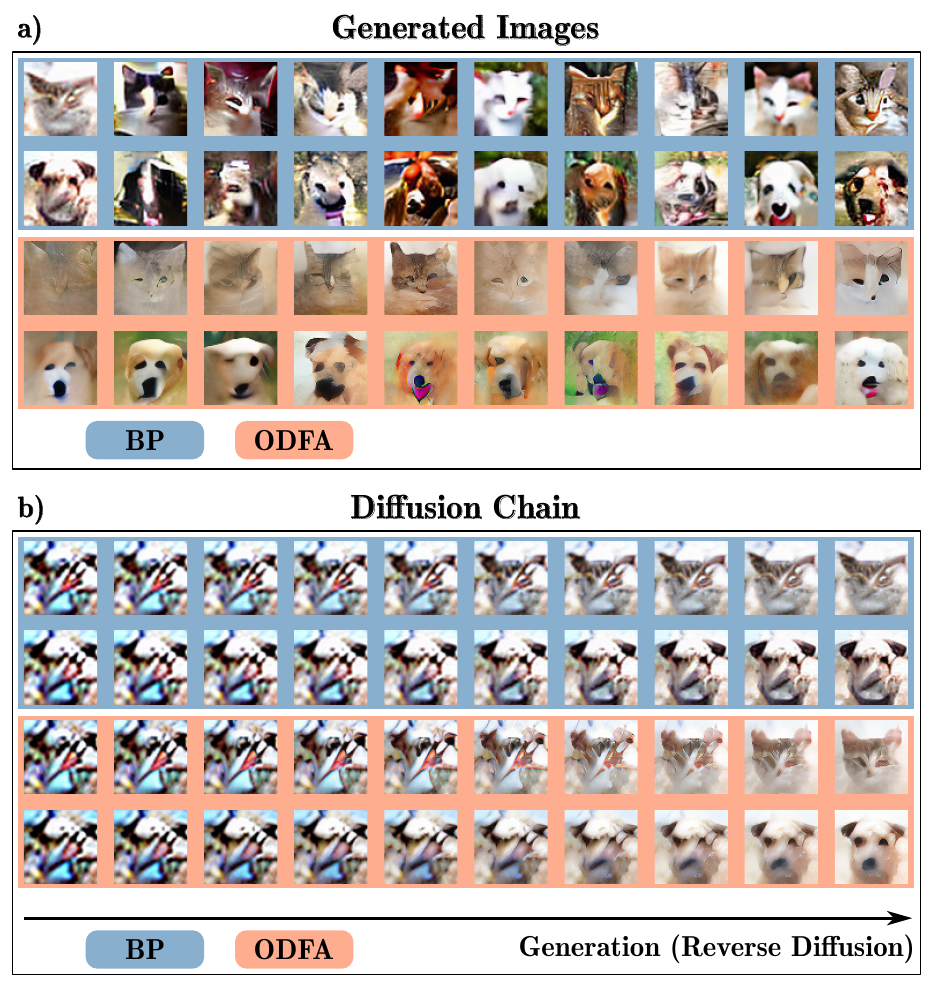}
  \caption{
  \textbf{Samples of DiT-B/2 trained by different methods on Animal Faces dataset.} Samples are generated with a classifier-free guidance scale of $3$, using $480$ DDPM steps and a ft-EMA VAE decoder. \textbf{a}, Generated samples. \textbf{b}, Reverse diffusion chains.
}
  \label{fig:ditafgenchain}
\end{figure*}

Beyond the numerical observations, we also present generated samples and reverse diffusion on AFHQ dataset through the DiT-B/2 trained by ODFA, as shown in Fig.\ref{fig:ditafgenchain}. The first panel displays samples produced by the DiT-B/2 using BP and ODFA. Though BP clearly produces better samples from the diffusion model's perspective, ODFA-trained DiT still generates visually coherent pictures of animal faces, with an overall realistic color. The performance gap between BP and ODFA can also be observed through the slightly more blurring around edges or mixed coloration in the ODFA-trained DiT. Despite these artifacts, the general identity and structure of each animal class remain recognizable. Figure~\ref{fig:ditafgenchain}(b) presents a series of reverse diffusion chains for specific cat and dog images, starting from the same noisy latent, using BP- or ODFA-trained DiT. The diffusion chain is given with a CFG scale of $3$ over $480$ DDPM steps. These continuous chains of the generative process again reveal that ODFA does not damage the essential denoising functionality; each latent still converges toward meaningful face details. Although BP’s chains appear more refined, the final images from ODFA remain plausible enough for most practical purposes. A noteworthy detail in these samples is the absence of background scenes in ODFA's outcomes. Even though BP's results visually seem not to be organized, they clearly separate the colors for the animal faces and for the background scenes. While for ODFA, most backgrounds share similar hues with the animal faces in the foreground. The good point is ODFA not being stuck at this phase, where we can notice some generated samples starting to form a background scene. In summary, shifting from digits on MNIST to the more colorful and detailed Animal Faces dataset demonstrates that ODFA’s performance extends beyond simplistic settings. The final images, although occasionally blurred, exhibit enough realism to confirm that ODFA can track the underlying diffusion process. Future work may refine ODFA's integration, potentially improving the performance. For now, these results reinforce ODFA’s applicability to diffusion models.

\section{Further details about scaling towards extreme-scale models} \label{sec:els}
In this section, we describe how we measured training times for both BP (GPU and CPU) and ODFA (GPU, CPU, and OPU), and explain how the DeepSpeed package broadens our training time measurements, revealing ODFA’s significant advantage over BP in this context.

\subsection{Training time measurement workflow with GPU and OPU} \label{sec:elsgpu}
\begin{figure*}[!htp]
  \centering
  \includegraphics[width=1.\linewidth]{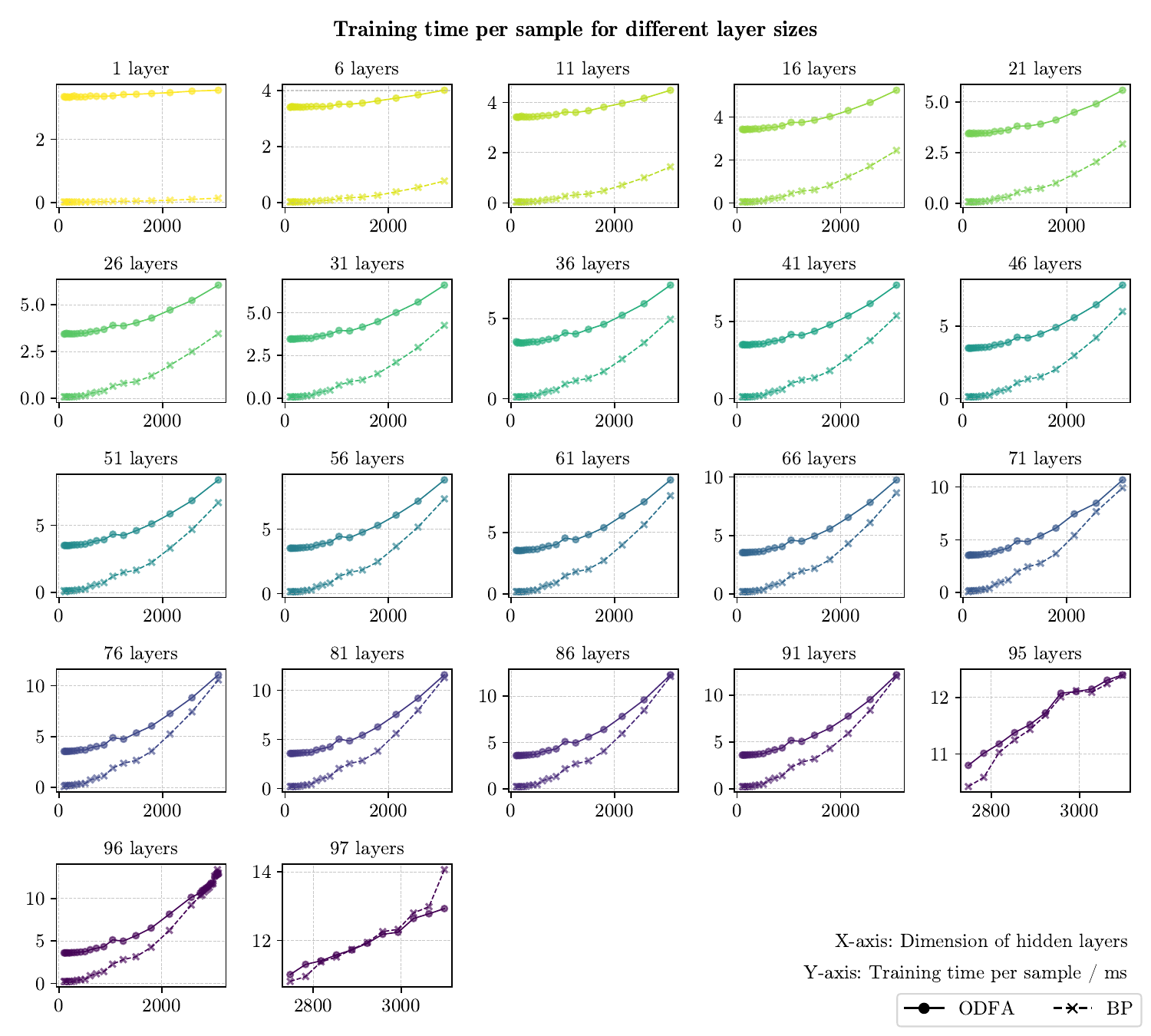}
  \caption{
  \textbf{Training time of FCNNs versus hidden layer size across different layer counts.} A decomposed figure from the main text Fig.~\ref{fig:scaling}(a). Each subplot depicts the training time per sample (in milliseconds) as a function of the hidden layer size ($100$ to $3080$ neurons) for a specific number of layers, with axis labels and legend placed at the bottom right. Solid-dot lines represent ODFA and cross-dashed lines represent BP. As layer depth grows, the gap in training time between BP and ODFA narrows, and in the final three subplots ODFA becomes faster than BP , indicating a crossover point. Across all subplots, ODFA exhibits a shallower slope of time increase with hidden dimension, suggesting a scalability advantage over BP for deeper and wider FCNNs.
}
  \label{fig:elsgpufull}
\end{figure*}

One crucial point of this project is to highlight ODFA’s training-time advantage, as shown in Fig.~\ref{fig:scaling} of the main text. Therefore, it's pivotal for us to be consistent and clear about what we call the training time. To avoid any misleading conclusions, we emphasize that we define the ODFA training time as the total elapsed time needed to complete a fixed number of training steps, rather than the time to reach a particular performance threshold or execute a specific number of GPU-based FLOPS. Notably, we do not employ the commonly used ``compute budget" paradigm adopted in the deep learning community for the following reasons.

Typically, when scaling deep learning models, both a compute budget and a time budget are considered. In conventional GPU-based systems, these two budgets are tightly coupled: the compute budget, measured in total FLOPs, can be directly translated into a corresponding time investment, given the GPU specifications, because GPUs fundamentally execute FLOPs. Within this framework, DFA has consistently been shown to underperform BP~\cite{filipovich2022scaling}, regardless of model scale. However, OPUs decouple the relationship between compute budget and time investment. As discussed in SI Note~\ref{sec:expspeed}, OPUs perform computations optically rather than through transistor-based FLOPs, rendering GPU-centric metrics such as FLOPs ambiguous or inapplicable for defining a ``compute budget" for our optical hardware. The lack of a unified metric consequently makes it challenging to conduct fair comparisons of ODFA and BP under a shared compute budget, though it also suggests that ODFA may deviate from previous conclusions regarding DFA versus BP. To bypass this ambiguity, we instead focus on the second budget, real-world time to finish a certain training, which remains hardware-agnostic and directly measurable.

The time we measure is the total elapsed wall-clock time required to complete the certain number of epochs in each case. Other than the raw dataset loading and the model creation, there is no other procedure not included in the measured training time. At the moment we start the timer, we synchronize the GPU; the OPU is engaged only after this point (no preprocessing on the OPU), and no further communication or control signals between the computer and the OPU occur once the timer stops. We train a standard FCNN (no LayerNorm or Dropout) on a generated random dataset of size $1$k. We apply an Adam optimizer with a fixed $1\times 10^{-3}$ learning rate and a batch size of $32$. The training runs for $20$ epochs, and before stopping the timer we ensure a full synchronization for GPU (BP) or both GPU and OPU (ODFA). Thus, all key processes, data transformation by the dataloaders, ternarized encoding, GPU/CPU/OPU communication delay, are included in the timing. For a sanity check, we also have a separate timer only to measure between epochs $5-15$, treating the first and last five epochs as warmup and cooldown, respectively. This separate timer gives consistent results.

In Fig.~\ref{fig:elsgpufull}, we provide a more detailed mesured training time data, of what was condensed into the main text Fig.~\ref{fig:scaling}(a). It displays how the training time of ODFA and BP changes as hidden layers grow denser. Within each subplot, the horizontal axis denotes the number of neurons in each hidden layer (ranging from $100$ to $3080$), while the vertical axis shows training time per sample in milliseconds. When examining these subplots, three major patterns emerge: 

\noindent1. As the number of layers increases, the time curves for ODFA and BP grow closer. In shallow networks, BP remains consistently faster, even closer to $0$ms. While ODFA always requires a nearly constant time even for the smallest network, due to the communication between the computer and the OPU, electro-optical signal conversion, etc. However, for networks above roughly $50$ layers, the gap narrows significantly, with ODFA’s overhead scaling less dramatically than BP’s. This phenomenon indicates that, while ODFA initially introduces extra complexity for smaller or shallower networks, it becomes more attractive at higher depths; 

\noindent2. In the subplots featuring $91$, $95$, $96$, and $97$ layers, there is a crossover event where ODFA’s training time actually shorter than BP’s. Concretely, for a $96$-layer FCNN with $3080$ neurons in each hidden layer, ODFA requires $13.09$ ms per sample, whereas BP requires $13.39$ ms. Although this difference appears small in absolute terms, it demonstrates a critical threshold at which optical feedback becomes more time-efficient than BP under this context. The shift occurs precisely where the model depth and width become extremely large, suggesting ODFA becomes advantageous at scale;

\noindent3. Most importantly, across all plots, ODFA’s curve always exhibit a better slope than BP’s as hidden dimensions grow. Even in cases where BP remains faster overall, once layer dimensions exceed $\sim1000-2000$ neurons, BP’s training time escalates more rapidly. ODFA, by contrast, displays a more tempered growth in training time as these dimensions expand. This pattern implies that while ODFA may begin slower in smaller networks, it gains rapidly once the model size pushes into the extreme-large regime.

\subsection{Extra-large scaling with memory offloading} \label{sec:elsol}

\begin{figure*}[!htp]
  \centering
  \includegraphics[width=1.\linewidth]{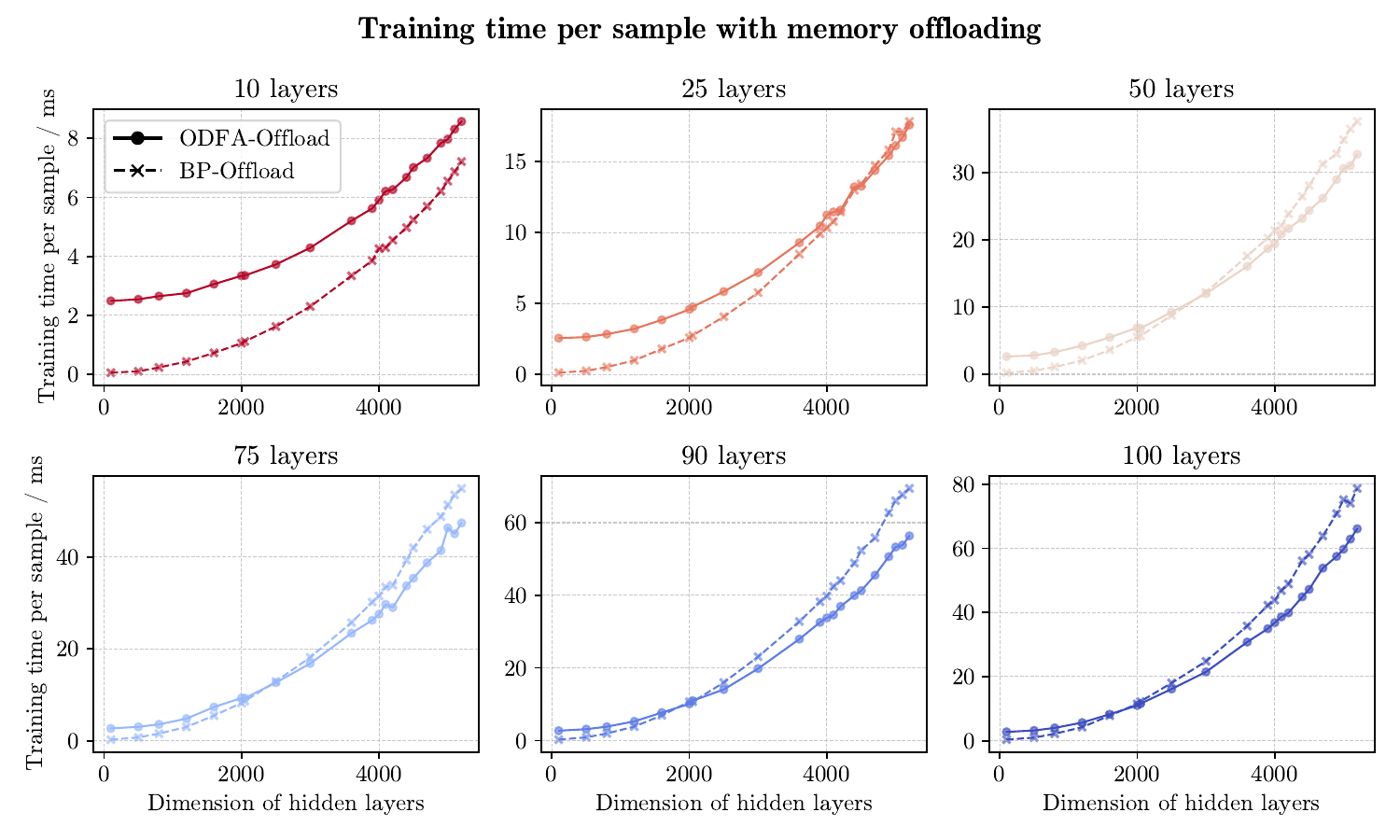}
  \caption{
  \textbf{Training time with memory offload of increasing depth and width.} Each subplot is decomposed from the main text Fig.~\ref{fig:scaling}(c), illustrating the training time per sample for FCNNs (up to $2.7$B parameters) of different layer counts and hidden layer sizes (up to $5200$ neurons), comparing ODFA and BP. A standard offloading technique from DeepSpeed~\cite{rasley2020deepspeed} is employed, transferring optimizer states and gradients to CPU RAM after each GPU-based gradient calculation (BP) or OPU-GPU-based optical feedback (ODFA). Both methods suffer a speed penalty from CPU-GPU data transfers, and ODFA has no less additional steps or shorter time penalty due to the memory offload comparing with BP. Moreoever, ODFA’s overhead can be slightly higher due to existing swapping demands between GPU/CPU/OPU. Crucially, memory offload occurs only after the primary GPU/OPU operation is complete, meaning the BP/ODFA mechanism is unaffected by offloading. ODFA’s speed advantage persists even at extra-large hidden dimensions and high layer counts. These results allow the demonstration of ODFA’s scalability and robustness under extra-large model sizes.
}
  \label{fig:elsolfull}
\end{figure*}

\begin{figure*}[!htp]
  \centering
  \includegraphics[width=1.\linewidth]{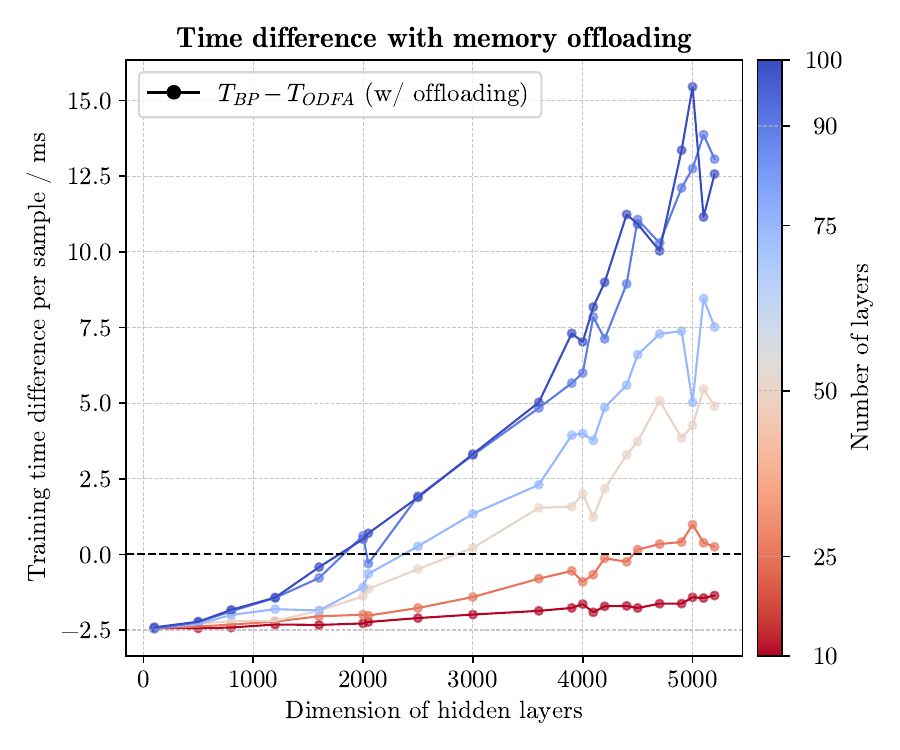}
  \caption{
  \textbf{Extended time difference with memory offload.} Under the same setting with the main text Fig.~\ref{fig:scaling}(c) and Fig.~\ref{fig:elsolfull}.
}
  \label{fig:elstmfull}
\end{figure*}

Even though the ODFA's training time is actually shorter than BP in Fig.~\ref{fig:elsgpufull}, the difference is not significant. Therefore, it's natural to question whether this advantage will continue and the scaling property remains the same, when the model size growing to the extra-large regime. To answer this question, we push our analysis toward extra-large models, but face a challenge when restricted to a single GPU-OPU with limited memory capacity. Consequently, we employ a widely-used memory offloading technique that moves optimizer states and gradients to CPU RAM following each training iteration. Because the Adam optimizer can require twice the storage size of the raw model weights, offloading these states provides a crucial workaround for memory bottlenecks but also harms overall training speed. We highlight that the gradient will only be moved to CPU after the optical feedback is delivered and the gradient propagation is done, thus ODFA does not inherently benefit from this approach. Actually, ODFA can sometimes face a slightly more serious penalty, due to ODFA’s additional CPU-OPU transfer may delay the offloading processes. This offloading approach, implemented through the DeepSpeed package~\cite{rasley2020deepspeed}, enables us to train extra-large models reaching up to $2.7$ billion parameters (a width of $5200$ neurons) on a single GPU, at the expense of frequent data transfers. 

Figure~\ref{fig:elsolfull} shows the per-sample training time across varying numbers of layers ($10$, $25$, $50$, $75$, $90$, $100$) and hidden layer sizes up to $5200$ neurons per layer, under a memory offloading. We present separate subplots for each layer depth, plotting the time in milliseconds on the vertical axis and the hidden dimension on the horizontal axis. In the main text Fig.~\ref{fig:scaling}(c) and here, we apply a standard DeepSpeed Offloading trick. The stage of ZeR0 Optimization is set to $2$, where the optimizer states and gradients can be moved to CPU RAM. The CPU memory is pinned for faster data transfer. We do not force the optimizer to be on CPU. Comparing these curves with those in prior figures (Fig.~\ref{fig:elsgpufull}), we have the inevitable slowdown introduced by data transfers between GPU and CPU. A observation is that both ODFA and BP suffer a penalty from offloading. The fundamental time cost of ODFA remains mostly unaffected. More detailed, the ODFA and BP occur during the gradient calculation step itself, whereas the memory offload is postponed until the GPU has already generated the necessary gradients. Consequently, the presence or absence of offloading does not directly affect the training time scaling. For clarity, Figure~\ref{fig:elstmfull} shows the time difference between ODFA and BP for these extra-large networks. We note that the initial time difference (here $\sim\!-2.5$~ms) is smaller than the one in the main text Fig.~\ref{fig:scaling}(a) ($\sim\!-3$~ms), but remains the similar.

Together, Figure\ref{fig:elsolfull} and Figure\ref{fig:elstmfull} conclude that ODFA’s training-time advantage is robust and can be more significant even in the extra-large regime, reinforcing the practical value of ODFA.

\subsection{Extended comparison of ODFA and DFA} \label{sec:elsdfa}

\begin{figure*}[!htp]
  \centering
  \includegraphics[width=0.8\linewidth]{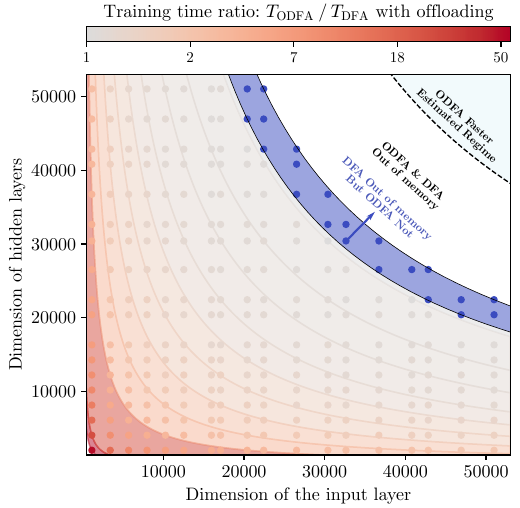}
  \caption{
  \textbf{ODFA vs. DFA training time ratio with offloading.} Under the same memory offload conditions and with a single hidden layer to enable larger-scale analysis, the ratio of training time per sample for ODFA relative to DFA ($T_{\text{ODFA}} / T_{\text{DFA}}$) decreases from $\sim\!50$ at low dimensions (consistent with non-offloading values of $\sim\!45$) to only about $2\%$ slower at the GPU’s memory limit. Beyond this limit, DFA cannot proceed while ODFA continues to train larger models until it reaches its own memory boundary. Notably, the gray-shaded region indicates the experimentally demonstrated scale ($45000\times45000$) where OPU-based random projections outperform GPU-based ones (see Fig.~3(a) of~\cite{ohana2020kernel}). Thus, while DFA is initially faster, ODFA offers superior scalability for training ultra-large models and can go beyond DFA's memory constraints.
}
  \label{fig:elsdfaol}
\end{figure*}

So far, ODFA has shown promise for large-scale model training, but its additional advantages over DFA beyond energy efficiency remain unclear. With the same offloading setting, Figure~\ref{fig:elsdfaol} provides an extended ODFA-DFA comparison of the per-sample training time ratio ($T_{\mathrm{ODFA}} / T_{\mathrm{DFA}}$), exploring larger dimensions than in the main text Fig.~\ref{fig:scaling}(d). In this test, we vary the input and hidden layer dimensions, both of which affect the speed of DFA and ODFA. We set the hidden layer count to $1$ to reach even larger scales. Because the layer count does not influence the relative time difference between DFA and ODFA, a large layer count will underestimate the true $T_{\mathrm{ODFA}} / T_{\mathrm{DFA}}$ ratio.

At smaller dimensions, the ratio of $T_{\mathrm{ODFA}} / T_{\mathrm{DFA}}$ is obviously very high ($\sim\!50$), which aligns with our non-offloading observations ($\sim\!45$). This outcome reflects the overhead that ODFA naturally carries. However, as the dimension increases and approaches the GPU’s memory boundary, the ratio decreases dramatically to near $1.02$. In other words, ODFA becomes only about $2\%$ slower than DFA at the largest model size that the GPU can still handle with DFA. Beyond this boundary, DFA cannot proceed because it hits a memory boundary. However, ODFA can continue to train larger models until it reaches its own memory boundary. This advantage arises because ODFA stores and processes the random feedback matrix in the optical domain. Another noteworthy point of Fig.~\ref{fig:elsdfaol} is the gray-shaded region, which marks the scale beyond $45000\times45000$. Random projections on OPU were previously shown to outperform GPU-based projections (see Fig.~3(a) of~\cite{ohana2020kernel}). 

This finding reinforces two key observations: 1. ODFA maintains a more favorable ``training time scaling" in the ultra-large regime; 2. ODFA can push model size beyond DFA under the same GPU memory constraints. 

\section*{Data availability}
The data that support the findings of this study are available from the corresponding author on reasonable request.

\end{document}